\documentclass[10pt,twocolumn,preprintnumbers,amsmath,amssymb,nofootinbib,superscriptaddress]{revtex4-1}

\usepackage{graphicx}
\usepackage{dcolumn}
\usepackage{amssymb,amsmath,bm}
\usepackage{color}
\usepackage[colorlinks,linkcolor=red,citecolor=blue,urlcolor=blue ]{hyperref}
\usepackage{multirow}
\usepackage{enumitem}
\usepackage{mathptmx}
\usepackage[utf8]{inputenc}

\newcommand{\nv}{\hat{\bf n}}
\newcommand{\resp}{{\cal A}}
\newcommand{\ligo}{LIGO}
\newcommand{\virgo}{Virgo}
\newcommand{\kagra}{KAGRA}
\newcommand{\lisa}{LISA}

\begin{document}
  \title{The $N_\ell$ of gravitational wave background experiments}
  \author{David Alonso}
  \thanks{Author list is alphabetised.}
  \email{david.alonso@physics.ox.ac.uk}
  \affiliation{Astrophysics, University of Oxford, DWB, Keble Road, Oxford OX1 3RH, UK}
  \author{Carlo R. Contaldi}
  \email{c.contaldi@imperial.ac.uk}
  \affiliation{Blackett Laboratory, Imperial College London, South Kensington Campus, London, SW7 2AZ, UK}
  \author{Giulia Cusin}
  \email{giulia.cusin@physics.ox.ac.uk}
  \affiliation{Astrophysics, University of Oxford, DWB, Keble Road, Oxford OX1 3RH, UK}
  \affiliation{Universit\'e de Gen\`eve, D\'epartement de Physique Th\'eorique and Centre for Astroparticle Physics, 24 quai Ernest-Ansermet, CH-1211 Gen\`eve 4, Switzerland}
  \author{Pedro G. Ferreira}
  \email{pedro.ferreira@physics.ox.ac.uk}
  \affiliation{Astrophysics, University of Oxford, DWB, Keble Road, Oxford OX1 3RH, UK}
  \author{Arianna I. Renzini}
  \email{arianna.renzini15@imperial.ac.uk}
  \affiliation{Blackett Laboratory, Imperial College London, South Kensington Campus, London, SW7 2AZ, UK}
  \date{Received \today; published -- 00, 0000}

  \begin{abstract}
    We construct a model for the angular power spectrum of the instrumental noise in interferometer networks mapping gravitational wave backgrounds (GWBs) as a function of detector noise properties, network configuration and scan strategy. We use the model to calculate the noise power spectrum for current and future ground-based experiments, as well as for planned space missions. We present our results in a language similar to that used in cosmic microwave background and intensity mapping experiments, and connect the formalism with the sensitivity curves that are common lore in GWB analyses. Our formalism is implemented in a lightweight {\tt python} module that we make publicly available at \url{https://github.com/damonge/schNell}.
  \end{abstract}

  \maketitle

  \section{Introduction}\label{sec:intro}
    The Universe is permeated by a faint background of gravitational waves fuelled by different mechanisms and generated at different epochs in history \cite{Caprini:2018mtu,2011RAA....11..369R}. A hypothesized period of accelerated expansion in the very early Universe will have amplified quantum fluctuations in the metric to macroscopic scales. Phase transitions will have stirred up the cosmic plasma and seeded metric fluctuations, through, for example, the collision of bubbles of different phases or through the dynamics and collapse of topological defects. The nonlinear dynamics of cosmic fields such as primordial magnetic fields will have generated metric perturbations, some of them in the form of gravitational waves. More recently, the inspiral and merger of binary systems, and the cataclysmic collapse of massive objects through supernovae or gamma ray bursts, will also make a substantial contribution to the gravitational wave background (GWB). See e.g. for black hole and neutron star mergers~\cite{TheLIGOScientific:2016wyq, Regimbau:2016ike, Mandic:2016lcn, Dvorkin:2016okx, Nakazato:2016nkj, Dvorkin:2016wac, Evangelista:2014oba}, \cite{Kelley:2017lek} for supermassive black holes, for exploding supernovae~\cite{Crocker:2015taa}, neutron stars~\cite{Surace:2015ppq, Talukder:2014eba, Lasky:2013jfa},  and stellar core collapse~\cite{Crocker:2017agi}, population III binaries~\cite{Kowalska:2012ba}.

    Measurements of the GWB are actively underway. At very low frequencies and large scales, the focus is on the cosmic microwave background (CMB). Primordial gravitational waves will distort the divergence-free component (the ``$B$-mode'') of the CMB polarization leading to a very distinct signature which can, hopefully, be isolated by current and future CMB experiments \citep{Seljak:1996gy,Kamionkowski:1996zd}. At higher frequencies, one hopes that the exquisite timing of multiple millisecond pulsars will pin down the presence of gravitational waves in the Galaxy \cite{Hobbs:2008yn,Mingarelli:2013dsa}. 

    In this paper we will concern ourselves with what we may be able to learn in the Hz and mHz range of wavelengths from gravitational wave interferometers, such as the Laser Interferometer Gravitational-Wave Observatory \ligo{} \citep{TheLIGOScientific:2016agk}, \virgo{} \citep{Accadia:2011zzc} and their successors from the ground, or space missions such as the Laser Interferometer Space Antenna \lisa{} \citep{Amaro2017,Ricciardone:2016ddg}. While the primary focus of these instruments is to detect and characterize individual events, they can also be deployed to scan the sky for the GWB. In this way, they are very much like other astrophysical survey instruments that are used to measure diffuse backgrounds. Two notable examples are the CMB experiments (already mentioned above) and intensity mapping experiments at radio frequencies which are used to look for integrated HI emission from radio galaxies \cite{Battye:2004re,Bull:2014rha}.

    The similarity between the quest for the GWB with interferometric experiments and other mapping experiments in cosmology and astrophysics is useful and has led to interesting developments. In particular, the application of techniques, first developed for the polarization of the CMB, to GWB has been fruitful \cite{Romano:2015uma,Gair:2014rwa}, being used to clarify some conceptual ambiguities \cite{Conneely:2018wis}, to develop a practical method for constructing GWB maps \cite{Renzini:2018vkx} and applied to the \ligo{} O1 and O2 runs \cite{Renzini:2019vmt}. A thorough and comprehensive description of multiple aspects of GWB detection can be found in \cite{Romano:2016dpx}.

    A key aspect in mapping diffuse backgrounds is the noise properties of the apparatus and how they are mapped onto the sky. A rough and often overly naive assumption is that the noise is ``white'', i.e. uncorrelated and homogeneous between pixels. Clearly this is not the case for a number of reasons: the noise in the time domain has a marked spectral dependence, the scan strategy will not cover the sky uniformly and, of particular relevance to the case we will study here, the envisaged networks of interferometers will only be able to reconstruct the largest angular scales. To have a reasonable characterization of the map-level noise properties in these experiments, a number of important effects must be included: the frequency (and potentially time) dependence of the detector noise, the instrumental response to a given sky signal, the spin of the quantity being mapped, and the geometric configuration of the network of antennas mapping the sky.

    In this paper we will attempt to do so by constructing a semi-analytic model for the angular power spectrum of the GWB instrumental noise (commonly labelled ``$N_\ell$''). The construction of this model will mimic in many ways how this is achieved for CMB and other diffuse mapping experiments and can be used for forecasting what one might expect with current and future GWB observatories. The approach we use to characterize the stochastic background signal projected onto a 2D sphere corresponds to the \emph{spherical harmonic decomposition method} employed in present searches at LIGO-Virgo, see e.g. \cite{Romano:2016dpx} for a recent review on the topic. We will not include, in the derivation of our instrumental noise model, the shot noise that arises from discrete GW sources, a topic which has been discussed elsewhere \cite{Jenkins:2019uzp, Cusinnew, Jenkins:2019nks, Alonso:2020mva}. Nevertheless, it must be included in the noise budget if one wishes to produce accurate forecasts of the signal to noise for planned observations.
    
    Much of the formalism and notation used in what follows is based on previous work. In particular, in \cite{Cornish:2001hg}, the general deconvolution problem of the sky signal, detector pattern and scan strategy was presented for LIGO and LISA, in \cite{2009PhRvD..80l2002T}, a maximum-likelihood estimator (and its corresponding covariance matrix) was applied to a multi-baseline array of detectors, in \cite{Romano:2015uma,Romano:2016dpx} the phase coherent mapping  of the gravitational wave sky was analyzed in detail, with a focus on the recovery of both divergence-free and curl-free modes of gravitational waves, and in \cite{Renzini:2018vkx} a complete, pixel based, map making algorithm was proposed for recovering the gravitational wave background from an array of interferometric detectors. In this paper, we add to this body of knowledge by providing a fast estimator of the noise angular power spectrum for a generic network of interferometers that can be used to quantify the detectability of different GWB models.

    We structure this paper as follows. In Section \ref{sec:gws} we present the formalism that allows us to link the GWB sky to the data, i.e. a time series associated to pairs of detectors. We then build an optimal quadratic estimator for the GWB intensity map, and use the map-level covariance of the estimator to construct a model for the angular power spectrum of the instrumental noise, $N_\ell$. In Section \ref{sec:Examples} we use our model to calculate different map-level noise properties, including the $N_\ell$, for a range of experiments: \lisa{}, an extended ground-based network made from the advanced Laser Interferometer Gravitational wave Observatories (\ligo{}) and their combination with other currently operational ground-based experiments, and the Einstein Telescope (ET). In Section \ref{sec:Discussion} we summarize our results and discuss them in the context of forecasts of the scientific returns of future GWB experiments.\\

  \section{Constructing a map of the gravitational wave background}\label{sec:gws}
    \subsection{Preliminaries}\label{ssec:gws.pre}
      Let us consider a bath of gravitational waves, $h(t, {\bf x})$. We can use the plane-wave expansion \cite{Allen:1996gp} to write:
      \begin{equation}\label{eq:h1}
        h_{ij}(t,{\bf x}) = \sum_{p\in\{+,\times\}}\int df\int d\nv^2\,h_p(f,\nv)\,e^{i\,2\pi f(t-\nv\cdot{\bf x})}\,{\sf e}^p_{ij}(\nv)\,,
      \end{equation}
      where $p$ labels the polarization states ($p\in\{+,\times\}$), $f$ is the frequency, $\nv$ is the direction of propagation in the sky, given in polar coordinates by,
      \begin{eqnarray}
        \nv \equiv (\cos\varphi\,\sin\theta,\sin\varphi\,\sin\theta,\cos\theta)\,,
      \end{eqnarray}
      and ${\sf e}^p_{ij}$ are the polarization tensors, constructed as follows:
      \begin{align}
        \hat{\bf l}   &\equiv (\sin\varphi,-\cos\varphi,0)\,,\\
        \hat{\bf m}   &\equiv (\cos\varphi\,\cos\theta,\sin\varphi\,\cos\theta,-\sin\theta)\,,\\
        {\sf e}^+_{ij}      &\equiv l_il_j-m_im_j\,,\\
        {\sf e}^\times_{ij} &\equiv l_im_j+m_il_j\,.
      \end{align}

      The second-order moments of $h_p$ are given by
      \begin{equation}\label{eq:h_correlator}
        \langle h_p(f,\nv)h_{p'}^*(f',\nv')\rangle \equiv \frac{1}{2}\delta^D(f-f')\frac{\delta^D(\nv-\nv')}{4\pi} {\sf W}_{pp'}(f,\nv)\,,
      \end{equation}
      where $\delta^{D}(\cdots)$ is the Dirac delta function, and the matrix ${\sf W}$ is related to the gravitational Stokes parameters through \cite{Renzini:2018vkx,Smith:2019wny}
      \begin{equation}\label{eq:h_covar}
        {\sf W}(f,\nv)\equiv
        \left(
        \begin{array}{cc}
          I(f,\nv)+Q(f,\nv) & U(f,\nv)-iV(f,\nv) \\
          U(f,\nv)+iV(f,\nv) & I(f,\nv)-Q(f,\nv)
        \end{array}
        \right)\,.
      \end{equation}

      Before moving forward, it is worth clarifying the notation we will use in the rest of the paper. Vectors are written as boldface symbols (e.g. ${\bf x}$), with unit vectors carrying a ``hat'' (e.g. $\nv$). Matrices are written with a sans-serif font (e.g. ${\sf W}$). Denoting the typical frequency our observatory is sensitive to $f$, $\tau$ will be the time scale over which we can approximate the detector position to be constant in celestial coordinates, and $T$ will be the typical time scale on which Fourier transforms are computed around a given time $t$. We have $\tau\gg T\gg1/f$. The Fourier transform of a quantity $X(t)$ computed in the interval $T$ will be denoted by
      \begin{equation}
        X_T(t, f)=\int_{t-T/2}^{t+T/2}dt'\,e^{-2i\pi ft'}X(t').
      \end{equation}
      We will also consider two different averaging procedures:
      \begin{enumerate}
        \item As implied by Eq.\,(\ref{eq:h_correlator}), gravitational waves can be modelled as Gaussian random fields with a covariance given by ${\sf W}$. With this rationale, $\langle X\dots Y \rangle$ (as in Eq.\,(\ref{eq:h_correlator})) will denote the ensemble average of the stochastic quantities $X\dots Y$ for a given underlying covariance ${\sf W}$.
        \item The Stokes parameters of the GWB will be considered random fields themselves, and we will denote their ensemble averages as $\langle I_1...I_N\rangle_W$.
      \end{enumerate}
      
      To simplify the notation, in what follows we will assume that the background is unpolarized ($Q=U=V=0$). We will also assume that the scale and frequency dependence of the background is factorisable, $I(f,\nv)=I_0(\nv){\cal E}_f$, where ${\cal E}_f$ is dimensionless and equal to 1 for a reference frequency $f_{\rm ref}$. It is common in the literature to express the GWB in terms of the fractional energy density in gravitational waves $\Omega_{\rm GW}$. This is related to the intensity through $\Omega_{\rm GW}(f,\nv)= (4\pi^2f^3)/(3H_0^2)I(f,\nv)$ \cite{Smith:2019wny}, where $H_0$ is the present value of the Hubble constant. We can therefore relate the power spectra of both quantities through
      \begin{align}\label{CellOmega}
        C^\Omega_\ell(f) &= \left(\frac{4\pi^2}{3H_0^2}f^3\,{\cal E}_f\right)^2 C_\ell^{I_0}.
      \end{align}
      In what follows, we will assume that intensity has a power-law frequency dependence, i.e.
      \begin{equation}\label{calE}
      {\cal E}_f=\left(\frac{f}{f_{\text{ref}}}\right)^{\alpha_I}\,,
      \end{equation}
       hence  $\Omega_{\rm GW}(f,\nv)\propto (f/f_{\text{ref}})^{\alpha}$ with $\alpha=3+\alpha_I$. When working out noise curves for different networks and scan strategies, we will focus on a fiducial spectrum  with $\alpha_I=-7/3$, corresponding to an astrophysical background of binary mergers. Unless otherwise stated, all power spectra and intensity maps shown here will be expressed in terms of $\Omega_{\rm GW}$, and not $I_0$.

    \subsection{Detector response}\label{ssec:gws.det}
      The response of a detector $A$ at position ${\bf x}_A$ to the gravity wave in Eq.\,(\ref{eq:h1}) can in general be written as
      \begin{align}\nonumber
        d_{A,T}(t,f)
        &= \int df'\delta_T(t,f-f')\int d\nv^2\,\sum_pF^p_A(f',\nv)h_p(f',\nv)\\\nonumber
        &\hspace{10pt}+n_{A,T}(t,f)\\\label{eq:response}
        &\simeq\int d\nv^2\,\sum_pF^p_A(f,\nv)h_p(f,\nv)+n_{A,T}(t,f).
      \end{align}
      Here $n_A(f)$ is a detector noise component, and we have defined
      \begin{equation}
        F^p_A(f,\nv)={\sf a}^{ij}_A\,{\sf e}^p_{ij}\,e^{-i2\pi f'\nv\cdot{\bf x}_A},
      \end{equation}
      where ${\sf a}_A(f, \nv, {\bf x}_A)$ is the detector response tensor, which we will specify in Section \ref{sec:Examples} for different cases. $\delta_T(t,f-f')$ is defined as
      \begin{align}\nonumber
        \delta_T(t,f-f')&\equiv\int_{-T/2}^{T/2}dt'e^{2\pi i(f-f')t'}\\
                        &\simeq \delta^D(f-f'),
      \end{align}
      where in the last line (as in the last line in Eq.\,(\ref{eq:response})) we have taken the limit $T\gg1/f$.

      In what follows we will use a discretized notation. The frequency range is divided into intervals of width $\Delta f=1/T$, where $T$ is the observation time of a given timeframe, and the celestial sphere is discretized into pixels labelled by an index $\theta$ with area $\Delta\Omega$. Dirac delta functions $\delta^D$ will get replaced by Kronecker deltas:
      \begin{equation}
        \delta^D(f-f')\rightarrow\frac{\delta^K_{ff'}}{\Delta f},\hspace{12pt}
        \delta^D(\nv-\nv')\rightarrow\frac{\delta^K_{\theta\theta'}}{\Delta\Omega}.
      \end{equation}
      The discretized version of Eq.\,(\ref{eq:response}) is therefore
      \begin{equation}
        d_{A,f}=\sum_{p,\theta}\Delta\Omega\,F^p_{A,f\theta}h_{pf\theta}+n_{A,f}.
      \end{equation}
      In what follows, it will be useful to write the timestreams of an array of detectors $A\in\{1,...,N\}$ as a vector ${\bf d}_f=(d_{1,f},...,d_{N,f})$, and likewise for its noise component.

    \subsection{Quadratic estimator}\label{ssec:gws.qml}
      Since $I_0$ is proportional to the variance of $h_{ij}$, an optimal quadratic estimator can be built for it with the form
      \begin{equation}
        \tilde{I}_{0,\theta} = \sum_{A,B,f}{\bf d}^\dagger_{f}{\sf E}_{\theta f}{\bf d}_{f}-b_\theta,
      \end{equation}
      where ${\sf E}_{\theta f}$ and $b_\theta$ are free coefficients to be determined by minimizing the variance of the estimator and eliminating its bias \cite{Tegmark:1996qt}.

      To begin with, the ensemble average of the product of two timestreams is
      \begin{align}
        \langle {\bf d}_f{\bf d}_{f'}^\dagger\rangle
        &=\frac{1}{2}\frac{\delta_{ff'}}{\Delta f}\left[\sum_{\theta }{\sf B}_{f\theta}\,I_{0,\theta}+{\sf N}_f\right]\\
        &\equiv\frac{1}{2}\frac{\delta_{ff'}}{\Delta f}\left[{\sf S}_f+{\sf N}_f\right]\equiv\frac{1}{2}\frac{\delta_{ff'}}{\Delta f}{\sf C}_f\,,
      \end{align}
      where we have defined
      \begin{equation}
        {\sf B}^{AB}_{f\theta}\equiv\Delta\Omega\,{\cal E}_f\sum_p F^p_{A,f\theta}F^{p*}_{B,f\theta},
      \end{equation}
      the noise power spectral density
      \begin{equation}\label{eq:npsd}
        \langle {\bf n}_{f}{\bf n}_{f'}^\dagger\rangle\equiv\frac{1}{2}\frac{\delta^K_{ff'}}{\Delta f} {\sf N}_f,
      \end{equation}
      and the frequency covariance ${\sf C}_f$ (with a signal component ${\sf S}_f$). Note that Eq.\,(\ref{eq:npsd}) is only valid when the detector noise is stationary, an approximation we assume here.

      In what follows it will be convenient to write ${\sf B}^{AB}_{f\theta}$ as:
      \begin{equation}
        {\sf B}^{AB}_{f\theta}=\frac{2}{5}\Delta\Omega\,{\cal E}_f\,\resp^I_{AB,f\theta}
      \end{equation}
      where $\resp^I_{AB}$ is the antenna pattern for the pair $AB$, in general given by
      \begin{equation}\label{eq:antenna}
        \resp^W_{AB}(\nv,f)=\gamma^W_{AB}\,e^{-i2\pi f\nv\cdot{\bf b}_{AB}}.
      \end{equation}
      $\gamma^W_{AB}$ are the overlap functions\footnote{Note that it is also common in the literature to include the complex baseline factor in Eq.\,(\ref{eq:antenna}) in the definition of $\gamma$. When focusing on measurements of the GWB monopole, the term ``overlap'' is also used to denote the sky average of these functions.}. Although we have only concerned ourselves with with the $I$ component of the Stokes parameters, they are in general given by
      \begin{align}\label{eq:gamma_I}
        \gamma^I_{AB}&\equiv\frac{5}{8\pi}\left[{\rm Tr}({\sf a}^T_A{\sf e}^+){\rm Tr}({\sf a}^T_B{\sf e}^+)^*+{\rm Tr}({\sf a}^T_A{\sf e}^\times){\rm Tr}({\sf a}^T_B{\sf e}^\times)^*\right]\nonumber\\
        \gamma^Q_{AB}&\equiv\frac{5}{8\pi}\left[{\rm Tr}({\sf a}^T_A{\sf e}^+){\rm Tr}({\sf a}^T_B{\sf e}^+)^*-{\rm Tr}({\sf a}^T_A{\sf e}^\times){\rm Tr}({\sf a}^T_B{\sf e}^\times)^*\right]\nonumber\\
        \gamma^U_{AB}&\equiv\frac{5}{8\pi}\left[{\rm Tr}({\sf a}^T_A{\sf e}^+){\rm Tr}({\sf a}^T_B{\sf e}^\times)^*+{\rm Tr}({\sf a}^T_A{\sf e}^\times){\rm Tr}({\sf a}^T_B{\sf e}^+)^*\right]\nonumber\\
        \gamma^V_{AB}&\equiv-i\frac{5}{8\pi}\left[{\rm Tr}({\sf a}^T_A{\sf e}^+){\rm Tr}({\sf a}^T_B{\sf e}^\times)^*-{\rm Tr}({\sf a}^T_A{\sf e}^\times){\rm Tr}({\sf a}^T_B{\sf e}^+)^*\right]\,.
      \end{align}
      The factor $5/8\pi$ is commonly introduced so that the overlap functions integrate to unity over the sphere for two co-located and perfectly aligned detectors with orthogonal arms.

      Assuming Gaussian statistics for $h_{pf\theta}$ and $n_{A,f}$, the mean and covariance of $\tilde{I}_\theta$ are
      \begin{align}\label{eq:mean_tilde}
        \langle \tilde{I}_{0,\theta}\rangle & =\sum_{\theta'} M_{\theta\theta'}I_{0,\theta'} + \tilde{b}_\theta - b_\theta,\\
        {\rm Cov}(\tilde{I}_{0,\theta},\tilde{I}_{0,\theta'})&=\frac{1}{2\Delta f^2}\sum_f{\rm Tr}\left({\sf E}_{\theta f}{\sf C}_f{\sf E}_{\theta' f}{\sf C}_f\right)
      \end{align}
      where we have defined the pixel-coupling matrix $M_{\theta\theta'}$ and the noise bias $\tilde{b}_\theta$:
      \begin{align}
        M_{\theta\theta'}\equiv\frac{1}{2\Delta f}\sum_f{\rm Tr}\left({\sf B}_{f\theta'}{\sf E}_{f\theta}\right),\hspace{12pt}
        \tilde{b}_\theta\equiv\frac{1}{2\Delta f}\sum_{f}{\rm Tr}\left({\sf E}_{\theta f}{\sf N}_f\right).
      \end{align}
      The bias term is therefore $b_\theta=\tilde{b}_\theta$, and the coefficients that minimize the variance are \cite{Tegmark:1996qt}
      \begin{equation}
        {\sf E}_{\theta f} = K {\sf C}^{-1}_f{\sf B}_{f\theta}{\sf C}^{-1}_f,
      \end{equation}
      where $K$ is an arbitrary constant\footnote{$K$ can be fixed, for example, by requiring $M_{\theta\theta}=1$, as done in \cite{Tegmark:1996qt}.}.

      Using this result, the pixel-coupling matrix, the noise bias, and the covariance of $\tilde{I}_\theta$ are
      \begin{align}
        &M_{\theta\theta'}=\frac{K}{2\Delta f}\sum_f{\rm Tr}\left({\sf C}^{-1}_f{\sf B}_{f\theta'}{\sf C}^{-1}_f{\sf B}_{f\theta}\right),\\
        &b_\theta=\frac{K}{2\Delta f}\sum_{f}{\rm Tr}\left({\sf C}^{-1}_f{\sf B}_{f\theta}{\sf C}^{-1}_f{\sf S}_f\right),\\
        &{\rm Cov}(\tilde{I}_{0,\theta},\tilde{I}_{0,\theta'})=\frac{K}{\Delta f}M_{\theta\theta'}.
      \end{align}

      Finally, as shown by Eq.\,(\ref{eq:mean_tilde}), a truly unbiased estimator for $I_{0,\theta}$ can be found by inverting $M$: $\hat{I}_{0,\theta}\equiv \sum_{\theta'}({\sf M}^{-1})_{\theta\theta'}\tilde{I}_{0,\theta'}$. The inverse covariance of $\hat{I}_{0,\theta}$ is therefore given by
      \begin{equation}\label{eq:invcov_discrete}
        {\rm Cov}^{-1}(\hat{I}_{0,\theta},\hat{I}_{0,\theta'})=\frac{1}{2}\sum_f{\rm Tr}\left({\sf C}^{-1}_f{\sf B}_{f\theta'}{\sf C}^{-1}_f{\sf B}_{f\theta}\right).
      \end{equation}
      A quadratic estimator for GWB intensity maps based on the full likelihood is implemented in \cite{AIRinprep}.

    \subsection{Fast estimate of the noise power spectrum}\label{ssec:gws.cont}
      We now use the results from the previous section to derive an expression for the angular noise power spectrum of a given GW experiment. We will assume we are in the noise-dominated regime, in which case ${\sf C}_f={\sf N}_f$.

      First, let us rewrite Eq.\,(\ref{eq:invcov_discrete}) as
      \begin{align}\nonumber
        \frac{{\rm Cov}^{-1}_{\theta\theta'}}{(\Delta\Omega)^2}=\frac{T}{2}\sum_{ABCD}\int df &\left(\frac{2{\cal E}_f}{5}\right)^2({\sf N}^{-1}_f)^{AB}({\sf N}^{-1}_f)^{CD}\\\label{eq:invcov_timeframe}
        &\resp^I_{BC}(f,\nv)\resp^I_{DA}(f,\nv),
      \end{align}
      where we have used $\Delta f=1/T$ and taken the continuum limit in $f$. Equation \ref{eq:invcov_timeframe} corresponds to the inverse map covariance of the intensity map originating from a timeframe with period $T$. Integrating over several such timeframes, the final covariance is
      \begin{align}\nonumber
        \frac{{\rm Cov}^{-1}_{\theta\theta'}}{(\Delta\Omega)^2}=\int dt\sum_{ABCD}\frac{1}{2}\int df &\left(\frac{2{\cal E}_f}{5}\right)^2 ({\sf N}^{-1}_f)^{AB}({\sf N}^{-1}_f)^{CD}\\\label{eq:map_noise_covar}
        &\resp^I_{BC}(f,\nv)\resp^I_{DA}(f,\nv),
      \end{align}

      Equation\,(\ref{eq:map_noise_covar}) provides an expression for the pixel-pixel noise covariance of a given map of the GWB intensity. In order to transform this into an estimate of the effective harmonic-space noise power spectrum, we start by computing the signal-to-noise ratio of the intensity map as
      \begin{equation}
        \left(\frac{S}{N}\right)^2=\left\langle\sum_{\theta\theta'}\hat{I}_{0,\theta}\,{\rm Cov}^{-1}_{\theta\theta'}\hat{I}_{0,\theta'}\right\rangle.
      \end{equation}
      Expanding $I_0(\nv)$ in spherical harmonics $I_0(\nv)=\sum_{\ell m} a_{\ell m} Y_{\ell m}(\nv)$ and using the definition of the angular power spectrum $C_\ell$ that $\langle a_{\ell m}a^*_{\ell'm'}\rangle_W\equiv C_\ell\,\delta_{\ell\ell'}\delta_{mm'}$, we obtain
      \begin{align}\label{SNR}
        \left(\frac{S}{N}\right)^2=\sum_\ell (2\ell+1) \frac{C_\ell}{N_\ell}\,,
      \end{align}
      where we have defined the inverse noise power spectrum $N^{-1}_\ell$ as
      \begin{align}\label{eq:Inoise}
        N_\ell^{-1}\equiv\frac{1}{2}\sum_{ABCD}\int df\,\int dt\,G_\ell^{AB,CD}(t,f)\,,
      \end{align}
      and where
      \begin{align}\nonumber
        &G_\ell^{AB,CD}(t,f)\equiv\left(\frac{2 {\cal E}_f}{5}\right)^2\left({\sf N}^{-1}_f\right)^{AB}\left({\sf N}^{-1}_f\right)^{CD}\\
        &\hspace{63pt}\frac{\sum_m{\rm Re}\left(\resp^I_{BC,\ell m}(t,f)\resp^{I*}_{DA,\ell m}(t,f)\right)}{2\ell+1}\,,\\
        &\resp^I_{AB,\ell m}(t,f)\equiv\int d\nv^2 Y^*_{\ell m}(\nv)\,\resp^I_{AB}(t,f,\nv)\,.
      \end{align}
      These expressions are our main results and allow us to predict the angular noise power spectrum of a given experiment and scan strategy, Eq.\,(\ref{eq:Inoise}) which is a function a the reference frequency $f_{\text{ref}}$. Notice that the signal to noise (\ref{SNR}) for a given background component (i.e. for a given value $\alpha_I$ of (\ref{calE})) is frequency-independent by construction as signal and noise parts scale in the same way with the reference  frequency. A key assumption made in obtaining this result is that the time-integrated response of the instrument can be compressed into an angular power spectrum. In practice this will not be the case, the scan strategy cannot be designed to isotropise the noise over $m$-mode and the resulting estimator will be sub-optimal. This is a necessary assumption for any analytical estimate however, in practice, optimal estimators will have to account for inhomogeneous noise in $m$-modes. Full likelihood estimators~\cite{AIRinprep} achieve this by a full time integration of the inhomogeneous response. 

      \subsubsection{Rigid networks}\label{sssec:gws.cont.rigid}
        Eq.\,(\ref{eq:Inoise}) involves a two-dimensional integral over $t$ and $f$ where the integrand involves at least one computationally expensive spherical harmonic transform of the antenna pattern $\resp$. This can be simplified further for experiments with a constant configuration (e.g. constant arm lengths, angular separations between arms, relative detector positions). In this limit, the network changes as a function of time as a rotating solid, and the antenna patterns at different times are related to each other through simple three-dimensional rotations (i.e. $\resp(t,f,\nv)=\resp(f,{\sf R}_t\nv)$, where ${\sf R}_t$ is a time-dependent rotation matrix). This is a good decription for fixed ground-based detectors such as \ligo{}. Orbiting experiments such as \lisa{} experience small changes in configuration (e.g. ``breathing'' variations in arm length and angles~\cite{Nayak:2005kb}), but the ``solid-rotation'' approximation is still valid at first order (and adopted here).
        
        Since a rotation only mixes the $m$-modes for each fixed $\ell$ in $\resp_{\ell m}$, under the assumption that the detector noise properties are stationary, $G_\ell$ does not depend on $t$. In this case, the noise power spectrum is simply given by
        \begin{align}\label{eq:Inoise_stationary}
          N_\ell^{-1}\equiv\frac{T_{\rm obs}}{2}\sum_{ABCD}\int df\,G_\ell^{AB,CD}(f),
        \end{align}
        where $T_{\rm obs}$ is the total observing time. Of course, the noise will never truly be stationary in GW detectors, and this can have a strong impact on measurements (as discussed for example in~\cite{TheLIGOScientific:2016dpb}). 
      
      \subsubsection{Uncorrelated detectors}\label{sssec:gws.cont.uncorrelated}
        If the different detectors forming the array are uncorrelated, $({\sf N}^{-1}_f)^{AB}=\delta_{AB}/N^{A}_f$, where $N^A_f$ is the noise power spectral density (PSD) of detector $A$. In this case, the expressions for the noise power spectrum simplify further
        \begin{align}\label{eq:Inoise_uncorrelated}
          N_\ell^{-1}\equiv\frac{T_{\rm obs}}{2}\sum_{AB}\int df\,\left(\frac{2 {\cal E}_f}{5}\right)^2\frac{\sum_m\left|\resp^I_{AB,\ell m}(f)\right|^2}{N^A_fN^B_f(2\ell+1)}.
        \end{align}
        
        Note that Eq.\,(\ref{eq:Inoise_uncorrelated}) can be broken up into two terms, corresponding to the contributions from detector auto-correlations and cross-correlations:
        \begin{align}
          N_\ell^{-1}=&\frac{T_{\rm obs}}{2}\sum_A\int df\,\left(\frac{2 {\cal E}_f}{5N^A_f}\right)^2\frac{\sum_m\left|\resp^I_{AA,\ell m}(f)\right|^2}{(2\ell+1)}\\
          &+T_{\rm obs}\sum_{A,B>A}\int df\,\left(\frac{2 {\cal E}_f}{5}\right)^2\frac{\sum_m\left|\resp^I_{AB,\ell m}(f)\right|^2}{N^A_fN^B_f(2\ell+1)}.
        \end{align}
        It will often not be possible to model the detector noise properties accurately enough to reliably estimate the bias term of the quadratic estimator, arising solely from the use of auto-correlations. In that case, we can still map the GW intensity using only data from detector cross-correlations. This is the case, for instance, of the \ligo{} two-detector array and will apply also to the extended ground-based network considered in this paper. 
         In such cases, the noise power spectrum can be calculated by setting the first contribution in the previous equation to zero.
      
      \subsubsection{The monopole}\label{sssec:gws.cont.monopole}
        Let us now consider the case of an isotropic signal (i.e. $I_0(\nv)=\bar{I}_\nu$) in a narrow frequency interval (${\cal E}_{f,\nu}=\Delta f\delta(f-\nu)$). Again in the noise-dominated regime, the noise on the average signal $\bar{I}_\nu$ can be computed as:
        \begin{equation}\nonumber
          \frac{\sigma_\nu^{-2}}{\Delta\nu}=\frac{T_{\rm obs}}{2}\left(\frac{8\pi}{5}\right)^2\sum_{ABCD}\left({\sf N}^{-1}_\nu\right)^{AB}\bar{\gamma}_{BC}(\nu)\left({\sf N}^{-1}_\nu\right)^{CD}\bar{\gamma}_{DA}(\nu),
        \end{equation}
        where $\bar{\gamma}_{AB}(\nu)$ is the sky average of $\resp(\nu,\nv)$
        \begin{equation}
          \bar{\gamma}_{AB}(\nu)\equiv\int \frac{d\nv^2}{4\pi}\,\resp_{AB}(\nu,\nv).
        \end{equation}

        Using only the cross-correlation of two detectors, the equation above reduces to
        \begin{equation}
          \sigma_f^2=\frac{1}{\Delta f\,T_{\rm obs}}\left(\frac{5}{8\pi}\right)^2\frac{N^A_fN^B_f}{\bar{\gamma}_{AB}^2(f)},
        \end{equation}
        an expression that is commonly used in the GWB lore.

   \section{Examples: \lisa{}, the extended ground-based network, and the Einstein Telescope}\label{sec:Examples}
     We now proceed to model $N_\ell$ for a number of currently envisaged experiments. To do so, we need three main ingredients: the noise power spectral densities of the detectors, their response tensors (${\sf a}_A$ in Eq.\,\ref{eq:response}) and their motions as a function of time.
     
     In particular, we will consider the Laser Interferometer Space Antenna, an extended ground-based network, and the Einstein Telescope. Although the discussion will be carried out in reverse chronological order, this will allow us to explore the most general detector response first and then specialize the discussion for simpler detectors.
     \begin{figure}
       \begin{center}
         \includegraphics[width=0.47\textwidth]{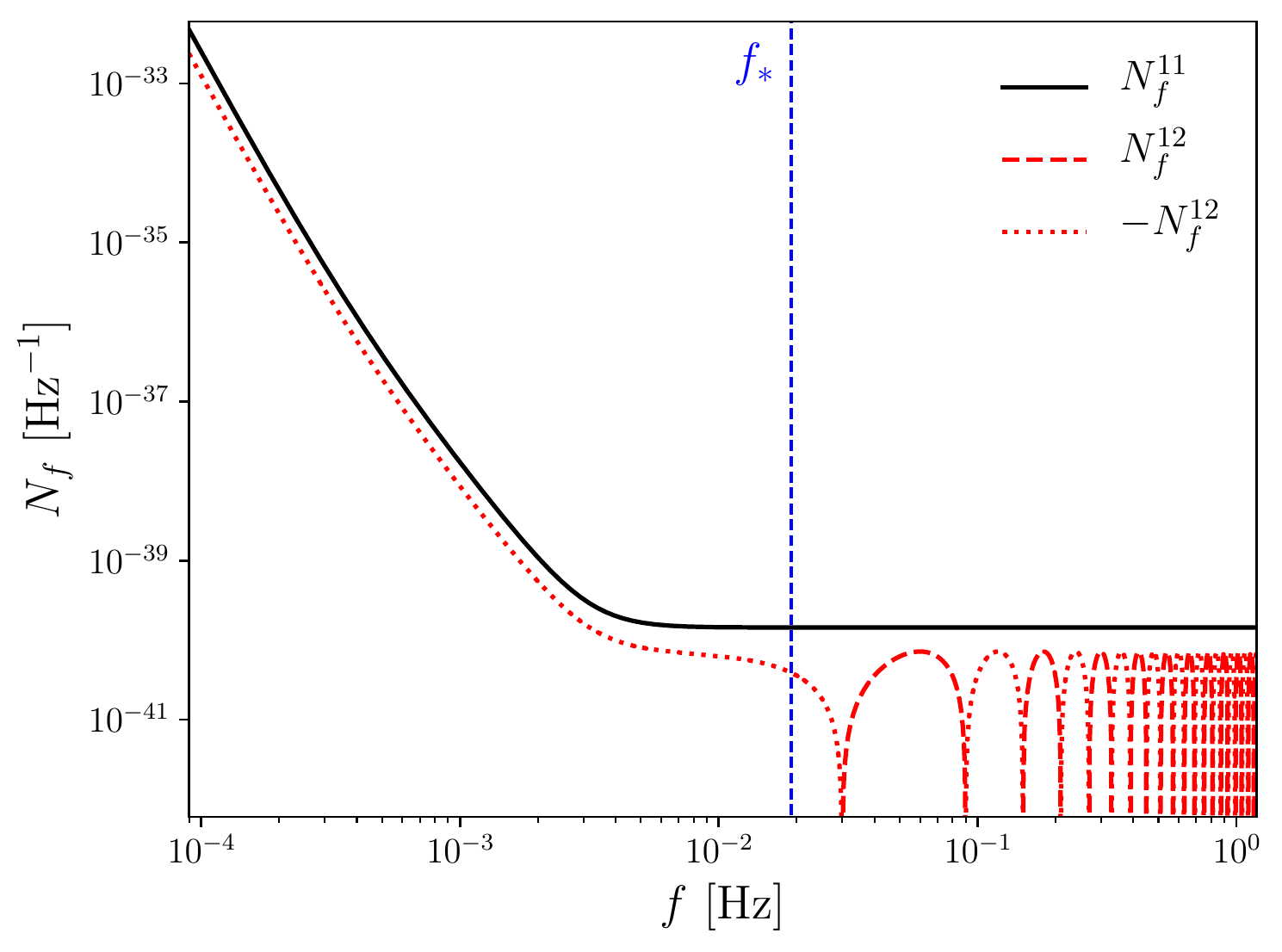}
         \caption{Power spectral density of the \lisa{} noise $N_f$. The noise variance (auto-correlation) is shown as solid black, while the cross-detector covariance is shown as dashed red, with its negative component in dotted red. Below the transfer frequency ($f_*$, shown as a vertical dashed blue line), the correlation coefficient (i.e. the ratio $N^{12}_f/N^{11}_f$) tends to $-1/2$.}\label{fig:LISA_psd}
       \end{center}
     \end{figure}

     \subsection{\lisa{}}\label{ssec:Examples.LISA}
         \begin{figure*}
         \begin{center}
             \includegraphics[width=0.7\textwidth]{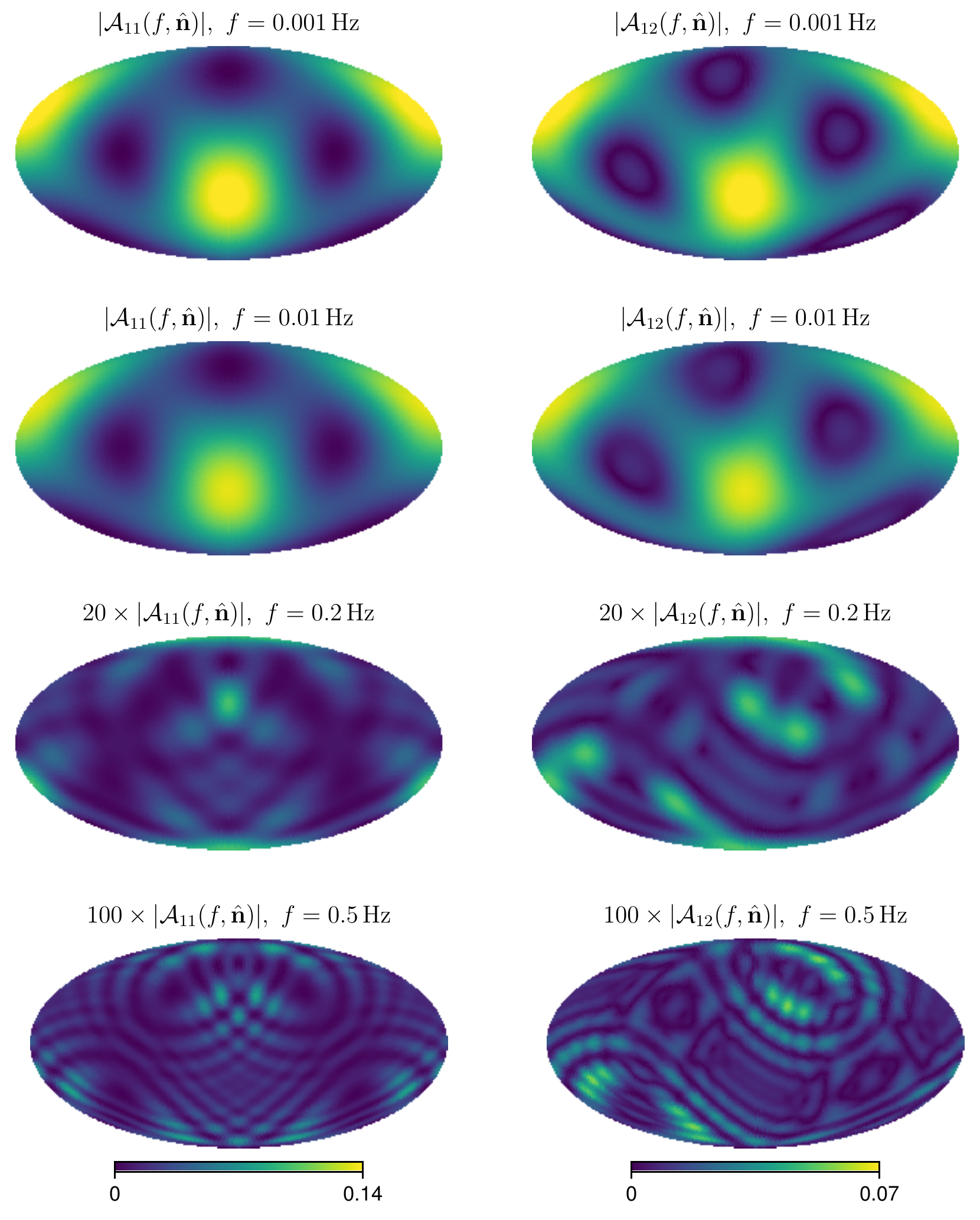}
             \caption{\lisa{} antenna pattern for auto- and cross-correlations between detectors (left and right panels respectively). Results are shown for different frequencies. A sharp transition in the antenna pattern, including a fast decrease in its amplitude, can be seen for $f>f_*\simeq0.02\,{\rm Hz}$.}
             \label{fig:LISA_antenna}
           \end{center}
         \end{figure*}
         \begin{figure*}
           \begin{center}
             \includegraphics[width=0.47\textwidth]{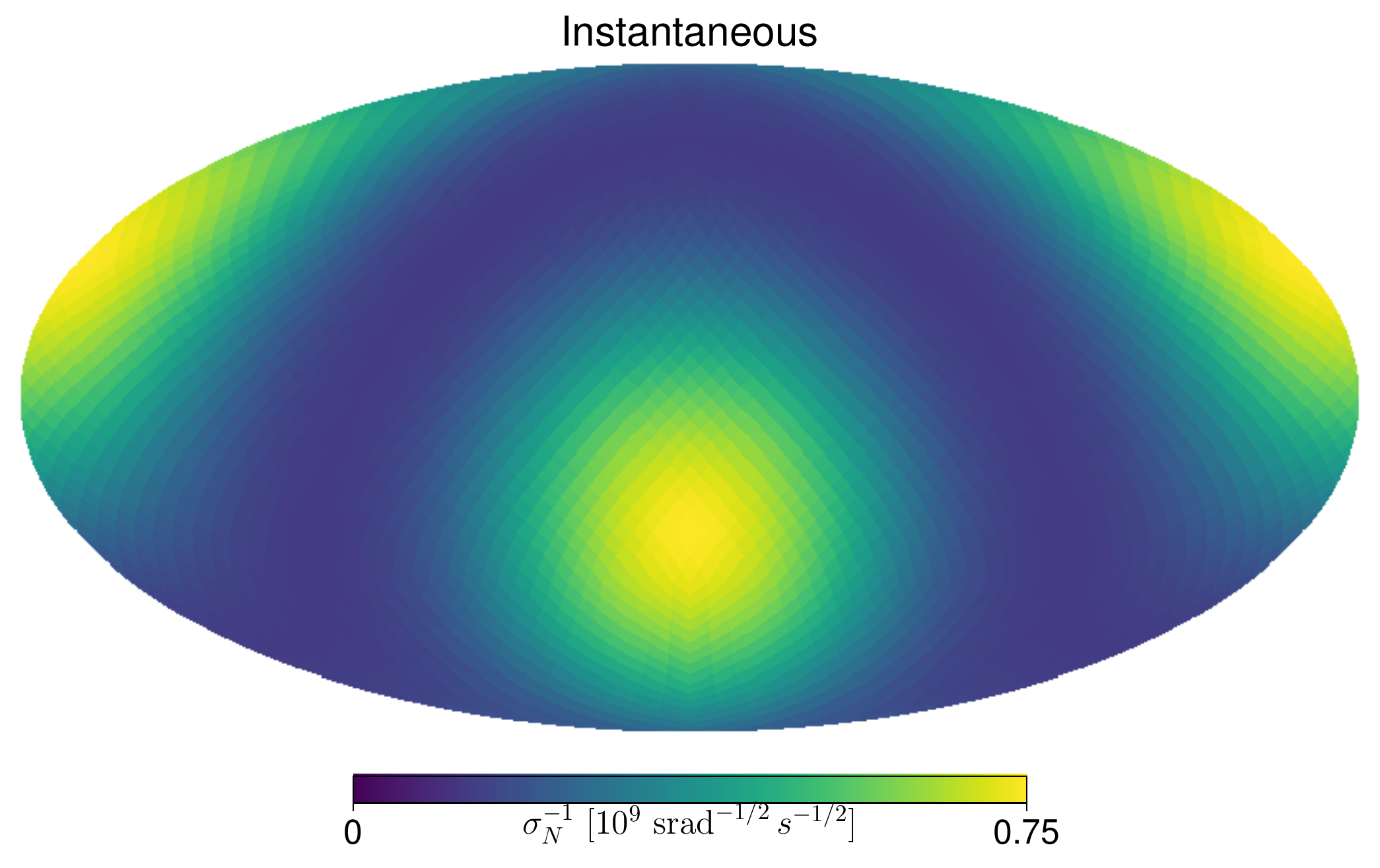}
             \includegraphics[width=0.47\textwidth]{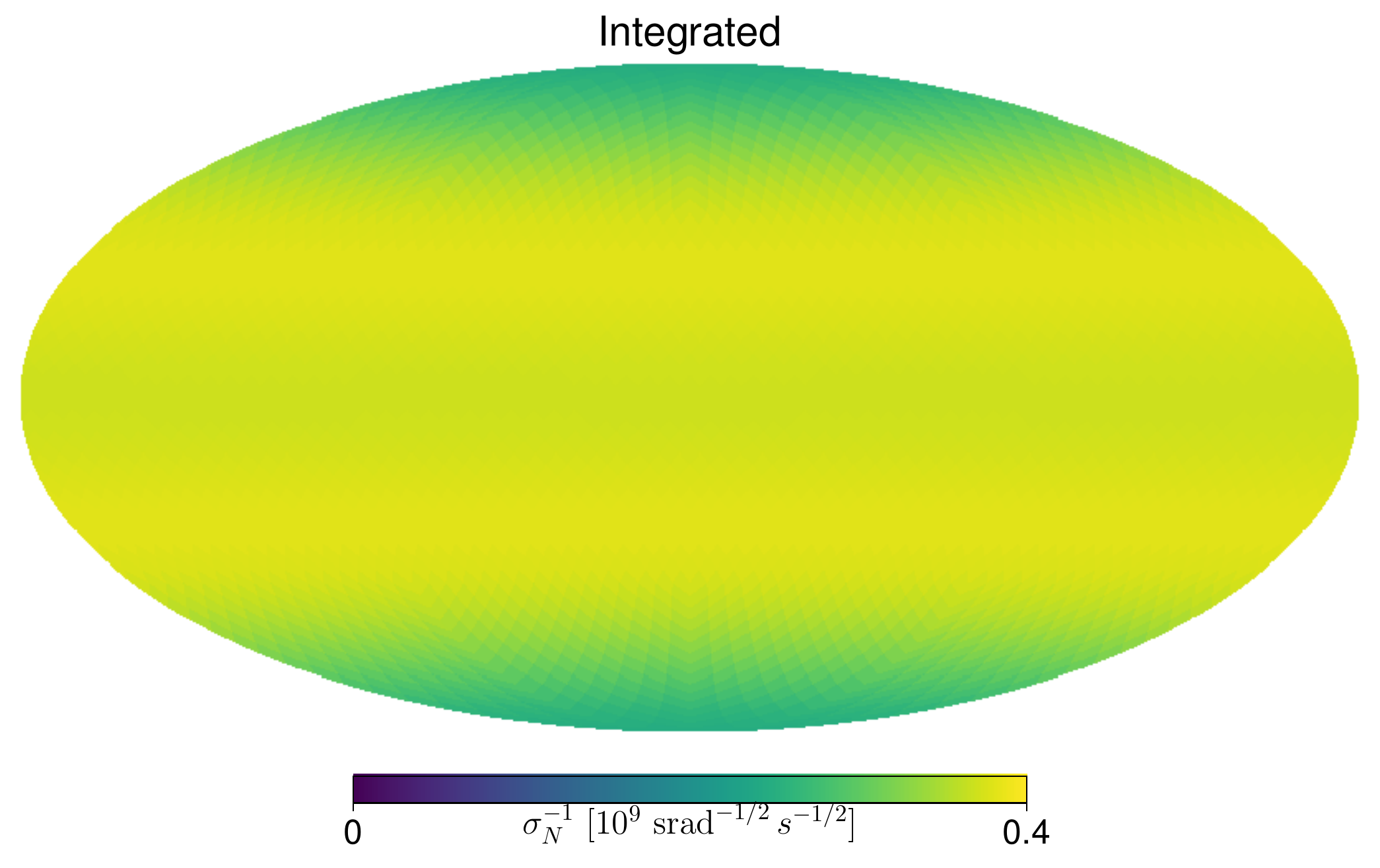}
             \caption{Square root of the map-level inverse noise variance for \lisa{} across the sky. Results are shown for a single timeframe (left panel) and for a full year of observation (right panel). The noise rms varies by less than 30\% across the sky after a full orbit.}
             \label{fig:LISA_noivar}
           \end{center}
         \end{figure*}
         \begin{figure}
           \begin{center}
             \includegraphics[width=0.47\textwidth]{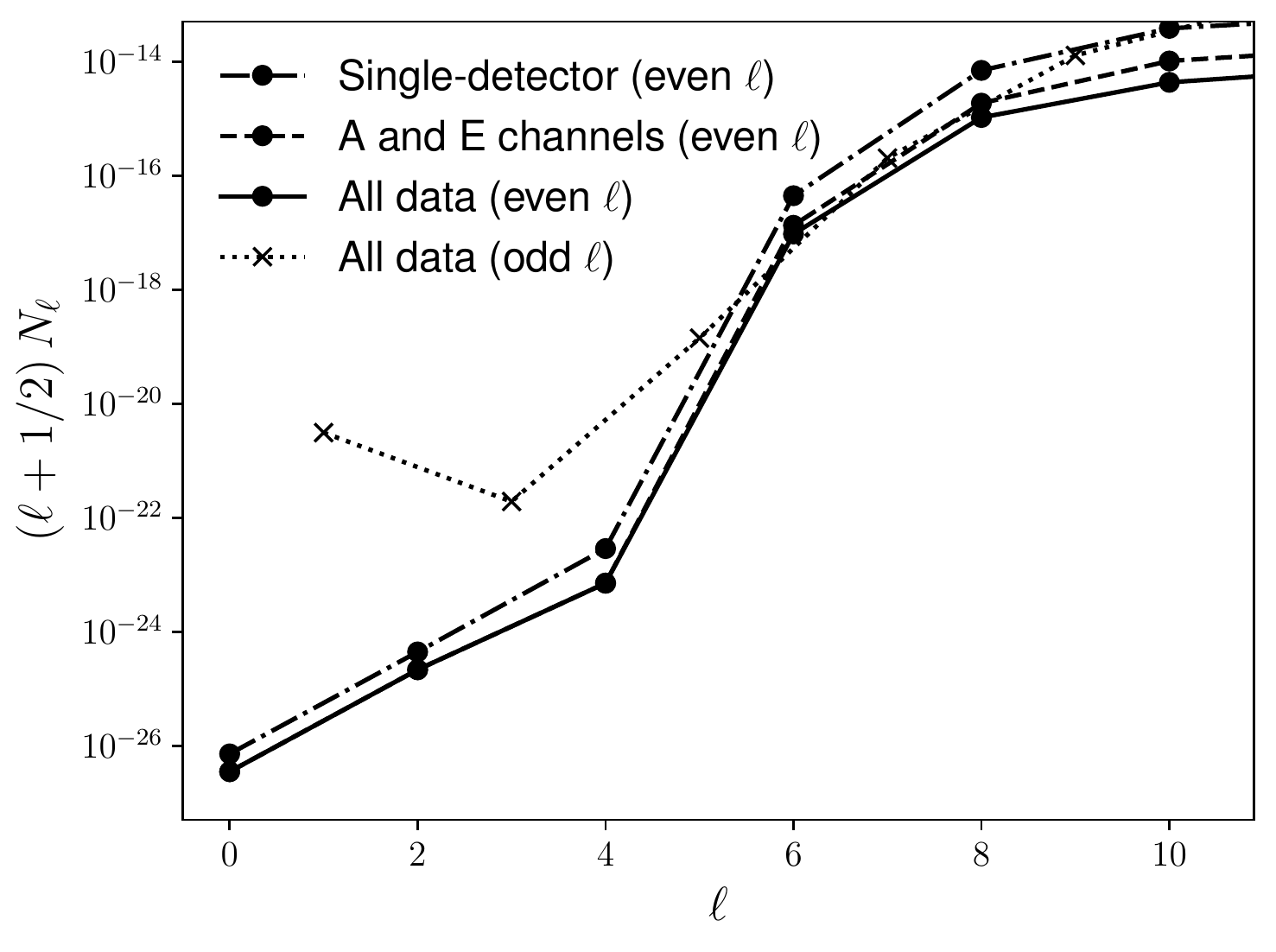}
             \caption{Angular noise power spectrum $N_\ell$ for different combinations of the \lisa{} detectors. The dot-dashed line shows the noise for a map constructed only from a single detector. Due to the parity of the antenna pattern for autocorrelations on low frequencies, odd $\ell$s cannot be recovered, and we show results only for even multipoles. The $N_\ell$ for the full \lisa{} network is shown in solid black, again for the even multipoles. The odd $\ell$s are shown as crosses connected by dotted lines. Although the addition of cross-correlations improves the noise for these modes, the sensitivity is still significantly poorer. Finally, the dashed line shows the $N_\ell$ for a map constructed only using the $A$ and $E$ \lisa{} channels, which dominate the overall sensitivity. Note that the expected level of the anisotropies for an astrophysical background is $C_\ell\sim10^{-30}$ at low $\ell$ \cite{Cusin:2019jhg}.}
             \label{fig:LISA_Nell}
           \end{center}
         \end{figure}
       The Laser Interferometer Space Antenna (\lisa{}, \citep{Amaro2017,Ricciardone:2016ddg}) is a space mission led by the European Space Agency, in collaboration with NASA, which is expected to fly in ten years time~\cite{Amaro2017}. The current design of the \lisa{} mission calls for three identical spacecraft flying in an equilateral triangular formation around the Sun, with arm length $L=2.5\cdot 10^9$ m. The center of mass of the detector (guiding center) is in a circular orbit at 1 AU and 20 degrees behind the Earth. In addition, the formation will also rotate in a retrograde motion with a one year period.

       \subsubsection{Detector motion}\label{sssec:Examples.LISA.motion}
         To describe the coordinates of the detector we work in a heliocentric, ecliptic coordinate system, following \cite{Rubbo_2004}. In this system the Sun is placed at the origin, the $x$-axis points in the direction of the vernal equinox, the $z$-axis is parallel to the orbital angular momentum vector of the Earth, and the $y$-axis is placed in the ecliptic to complete the right handed coordinate system. The individual \lisa{} spacecrafts will follow independent Keplerian orbits. The spacecraft positions as a function of time are derived e.g. in Appendix A of \cite{Rubbo_2004}, and Eq.\,(1) of this reference gives the cartesian coordinates as a function of time up to second order in the eccentricity.
         
         The arm lengths and angles undergo slow variations as a function of time (commonly called ``breathing''). We have verified that this leads only to $<0.1\%$ variations in the estimated noise angular power spectrum, and therefore all results presented here will assume a rigidly moving network of detectors.

       \subsubsection{Detector response}\label{ssec:Examples.LISA.response}
         The response tensor (introduced in Eq.\,(\ref{eq:response})) of a Michelson interferometer with two arms pointing along the unit vectors $\hat{\bf u}$ and $\hat{\bf v}$ to a gravitational wave propagating along the line of sight $\nv$ is given by \cite{Cornish_2001, 1975GReGr...6..439E, Rubbo_2004}
         \begin{equation}\label{a:LISA}
           {\sf a}^{ij}(f,\nv)\equiv\frac{1}{2}\left[u^iu^j{\cal T}(\nv\cdot\hat{\bf u},f)-v^iv^j{\cal T}(\nv\cdot\hat{\bf v},f)\right],
         \end{equation}
         where the transfer function ${\cal T}$ is
         \begin{align}\nonumber
           {\cal T}(\mu,f)=g(f)\,\frac{1}{2}&\left[{\rm sinc}\left(\frac{f}{2f_*}(1-\mu)\right)e^{-i\frac{f}{2f_*}(3+\mu)}\right.\\\label{eq:transfer}
           &\,\,\left.+{\rm sinc}\left(\frac{f}{2f_*}(1+\mu)\right)e^{-i\frac{f}{2f_*}(1+\mu)}\right].
         \end{align}
         Here $f_*\equiv c/(2\pi L)\simeq19\,{\rm mHz}$ is the {\sl transfer frequency}, corresponding to the frequency of a GW with a wavelength given by the arm length. $f_*$ marks a substantial change in the behaviour of the detector response. GWs with frequencies higher than $f_*$ undergo more than one oscillation within the detector arm, leading to self-cancellation effects that reduce the sensitivity to such waves. Below $f_*$, the transfer function approaches unity.
         
         The prefactor $g(f)$ depends on the time delay interferometry (TDI) combination used. For simplicity we will use the simplest TDI combination (referred to as TDI 1), for which $g(f)=1$. Our results, however, are insensitive to this choice, since the prefactor cancels out when using the noise PSD corresponding to the chosen TDI channel.

         There are two equivalent approaches to find Eq.\,(\ref{eq:transfer}). The first approach \citep{1975GReGr...6..439E, Rubbo_2004} is to find the Doppler shift of the photon emitted by the first spacecraft and received by the second. The second approach is to integrate along the photon's trajectory to find the path length variation caused by the gravitational wave \citep{Cornish_2001}. The two approaches give equivalent results for the response function at order $\mathcal{O}(vh)$ where $v$ is the spacecraft velocity.
     
       \subsubsection{Noise PSDs}\label{sssec:Examples.LISA.noise}
         The current  official model for the power spectral density of the \lisa{} noise $N_f$ is based on the Payload Description Document, and is referenced in the \emph{\lisa{} Strain Curves} document LISA-LCST-SGS-TN-001. The noise in the 3 \lisa{} spacecrafts is correlated, and therefore we need to characterize both the per-detector PSD and the correlation between detectors ($N^{AA}_f$ and $N^{AB}_f$ in the language of Section \ref{ssec:gws.qml}). For both, we use the fits provided in \cite{Smith:2019wny} (see also \cite{Robson_2019}). The two main sources of noise are the so-called ``acceleration noise'' and fluctuations in the optical path lengths between detectors (with the former dominating over the latter at low frequencies). Note that our calculation already accounts for the arm length penalty in the detector response for frequencies $f>f_*$, as well as the different normalization of the overlap functions for 60$^\circ$ arm apertures compared with that of orthogonal arms, and therefore we do not correct the noise PSDs to account for this (as is sometimes done in the literature). The resulting auto-correlation and cross-correlation PSDs are shown in Figure \ref{fig:LISA_psd}.

         It is common in the \lisa{} literature to make use of the linear combinations of the detector signals that diagonalise the noise covariance matrix. Our formalism automatically accounts for cross-detector correlations, and therefore we do not need to do so here, however, it will be instructive to explore the amount of information carried by each of these linear combinations. The normalized eigenvectors of the noise covariance give rise to the following uncorrelated linear combinations, commonly labelled the $A$, $E$ and $T$ channels:
         \begin{align}
           d_{A,f} &= \frac{1}{\sqrt{2}}(d_{1,f}-d_{3,f}),\\
           d_{E,f} &= \frac{1}{\sqrt{2}}(d_{1,f}-2d_{2,f}+d_{3,f}),\\
           d_{T,f} &= \frac{1}{\sqrt{3}}(d_{1,f}+d_{2,f}+d_{3,f}),
         \end{align}
         with their associated uncorrelated noise PSDs:
         \begin{equation}
          N^A_f=N^E_f=N^{11}_f-N^{12}_f,\hspace{12pt}N^T_f=N^{11}_f+2N^{12}_f,
         \end{equation}
         where $N^{11}_f$ and $N^{12}_f$ denote the diagonal and off-diagonal noise PSDs. As shown in Figure \ref{fig:LISA_psd}, the correlation coefficient of the noise PSDs tends to $-1/2$ at low frequencies, and therefore one would expect the $T$ channel to become noiseless in that limit. However, it is possible to show (see e.g. \cite{Smith:2019wny}) that the signal in that channel falls even faster with frequency, and that in fact most of the signal-to-noise is carried by the $A$ and $E$ channels. We will explore this in more detail in the next section.

       \subsubsection{Map noise properties}\label{sssec:Examples.LISA.map}
         With the various ingredients in hand we can now calculate the map-level noise properties. We will present results for a 4-year observation period with a reference frequency $f_{\rm ref}=0.01\,{\rm Hz}$. All maps are shown in Mollweide's projection in ecliptic coordinates.
         
         Figure \ref{fig:LISA_antenna} shows the antenna patterns for the auto-correlation of one of the \lisa{} detectors (left column) and for the cross correlation of two detectors (right column) at different frequencies. The antenna pattern changes significantly for frequencies $f<f_*$, decreasing fast in amplitude, due to the effect of the transfer function. The inverse noise variance across the sky for a single timeframe and for a full year observing time for \lisa{} is shown in Figure \ref{fig:LISA_noivar}. At any given time, the network is sensitive to signals coming from two antipodal directions, each covering an area of $\sim10,000\,{\rm deg}^2$. Within one full orbit, the network sweeps the sky, achieving a reasonably homogeneous coverage with $<30\%$ fluctuations in the noise standard deviation and an effective observed sky fraction of $f_{\rm sky}\sim0.97$.

         Figure \ref{fig:LISA_Nell} shows the resulting noise angular power spectrum for a map constructed from different linear combinations of the data from the three \lisa{} detectors. The noise power spectrum for a map constructed from a single detector is shown as a dot-dashed line. Due to the even parity of the antenna pattern ($\resp$ in Eq.\,(\ref{eq:antenna})) for detector auto-correlations (i.e. in the absence of the factor $\exp(-i2\pi f\nv\cdot{\bf b}_{AB})$), it is impossible to reconstruct the odd multipoles, and therefore we only show the noise power spectrum for even $\ell$. The solid black line shows the noise power spectrum using all the information in the three \lisa{} detectors. Data is again shown only for the even $\ell$, with the odd $\ell$s shown separately as crosses connected by a dotted line.
         
         Even in the presence of cross-detector correlations, there is a clear asymmetry between the sensitivity to even and odd $\ell$s. This is understandable: in order to break the parity of the antenna pattern and gain sensitivity to the odd $\ell$s, the factor $f\,\nv\cdot{\bf b}_{AB}/c$ should be of order $\sim1$. However, in the case of \lisa{}, the baseline between detectors $b_{\rm AB}$ is equal to the arm length $L$ and, as we described above, \lisa{}'s sensitivity decreases fast for wavelengths smaller than $L$. The only way to improve the noise for low odd multipoles would therefore be to include data from a second constellation of detectors separated from \lisa{} by a distance larger than $L$, for example as proposed for the Advanced Laser Interferometer Antenna (ALIA) project~\cite{Crowder2005}. 

         Finally, the dashed black line in Figure \ref{fig:LISA_Nell} shows the noise power spectrum for a map constructed only from the $A$ and $E$ \lisa{} channels (i.e. discarding all information from the $T$ channel). As described in the previous section, most of the information is carried by $A$ and $E$, and $T$ only becomes useful on scales $\ell\gtrsim6$.
         
         As shown in Fig. \ref{fig:LISA_Nell}, \lisa{} will only be sensitive to the largest scales, with $N_\ell$ increasing by more than two orders of magnitude between $\ell=0$ and $\ell=4$, and by an additional $\sim6$ orders of magnitude for $\ell=6$. Furthermore, our results indicate that the noise level achieved by \lisa{} will be above that expected for an astrophysical background, which would correspond to $C_\ell\sim10^{-30}$ at $\ell\sim2$ and $f_{\rm ref}=0.01\,{\rm Hz}$ \citep{Cusin:2019jhg}, making the detection of anisotropies challenging. Note that this does not apply to the monopole, sky-averaged signal, the amplitude of which is expected to be 4 to 5 orders of magnitude higher \citep{Chen:2018rzo, Cusin:2019jpv}.

     \subsection{Extended ground-based network} \label{ssec:Examples.aLIGO}
       There are currently four operating ground-based GW detectors at different locations around the Earth which probe the same frequencies. There are two \ligo{} detectors in the USA \citep{TheLIGOScientific:2016xzw}, one based in Hanford, Washington and one in Livingston, Louisiana; the \virgo{} detector based near Pisa, Italy \citep{Accadia:2011zzc}, and, most recently, the Kamioka Gravitational Wave Detector (\kagra{}), set in the Kamioka mine, Japan \citep{Aso:2013eba}. In what follows, we will consider a network made up of the two US detectors, \virgo{} and \kagra{}. The GEO600 detector based at Hannover, Germany \citep{Dooley:2015fpa} is not considered here as it is notably smaller than the others and is mainly used as a testing ground for new technologies. We will call the network made up of these four sites the ``extended ground-based network'' (XGN).

       \subsubsection{Network properties}\label{sssec:Examples.aLIGO.dets}
          \begin{table}
            \centering
            \begin{tabular}{| c | c | c | c |}
              \hline
              Name & Latitude (deg) & Longitude (deg) & $\xi$ (deg)  \\
              \hline
              Hanford     & 46.4 & -119.4 & 171.8 \\
              Livingston & 30.7 &  -90.8 & 243.0 \\
              \virgo{}       & 43.6 &   10.5 & 116.5 \\
              \kagra{}       & 36.3 &  137.2 & 225.0 \\
              \hline
              Einstein Telescope$^*$ & 40.1 & 9.0 & 90 \\
              \hline
            \end{tabular}
            \caption{Coordinates and orientation angles for the ground-based detectors considered here. The orientation angle $\xi$ is defined as the angle between the aperture bisector and the local parallel. See \cite{Seto:2008sr} for further details. The Einstein Telescope is different from the other experiments in several aspects. It is made up of three detectors in an equilateral triangle pattern. The coordinates given above are for the triangle's barycenter, and correspond to an arbitrary location in Sardinia, one of the possible sites for the experiment. The orientation angle is also arbitrary, and defined as the angle between the bisector of one of the vertices and the local parallel (i.e. the vertex is pointing north from the barycenter).} \label{tab:coords}
          \end{table}

          We model the network as an array of detectors with orthogonal arms located at fixed positions on the Earth surface. Each detector is defined by its latitude, longitude and orientation angle $\xi$ (i.e. the angle between the arm bisector and the local parallel) \cite{Seto:2008sr,Renzini:2018vkx}, as well as its noise power spectral density $N_f$. Table  \ref{tab:coords} lists the coordinates and orientations of all detectors considered here.

          In the case of these ground-based detectors, instrument noise limits the operational frequency well below their transfer frequency ($f_*\simeq12\,{\rm kHz}$ for a 4-km arm), and therefore their transfer function is 1. In this case, the response tensor is simply ${\sf a}^{ij}=(u^iu^j-v^iv^j)/2$. 

          \begin{figure}
            \begin{center}
              \includegraphics[width=0.47\textwidth]{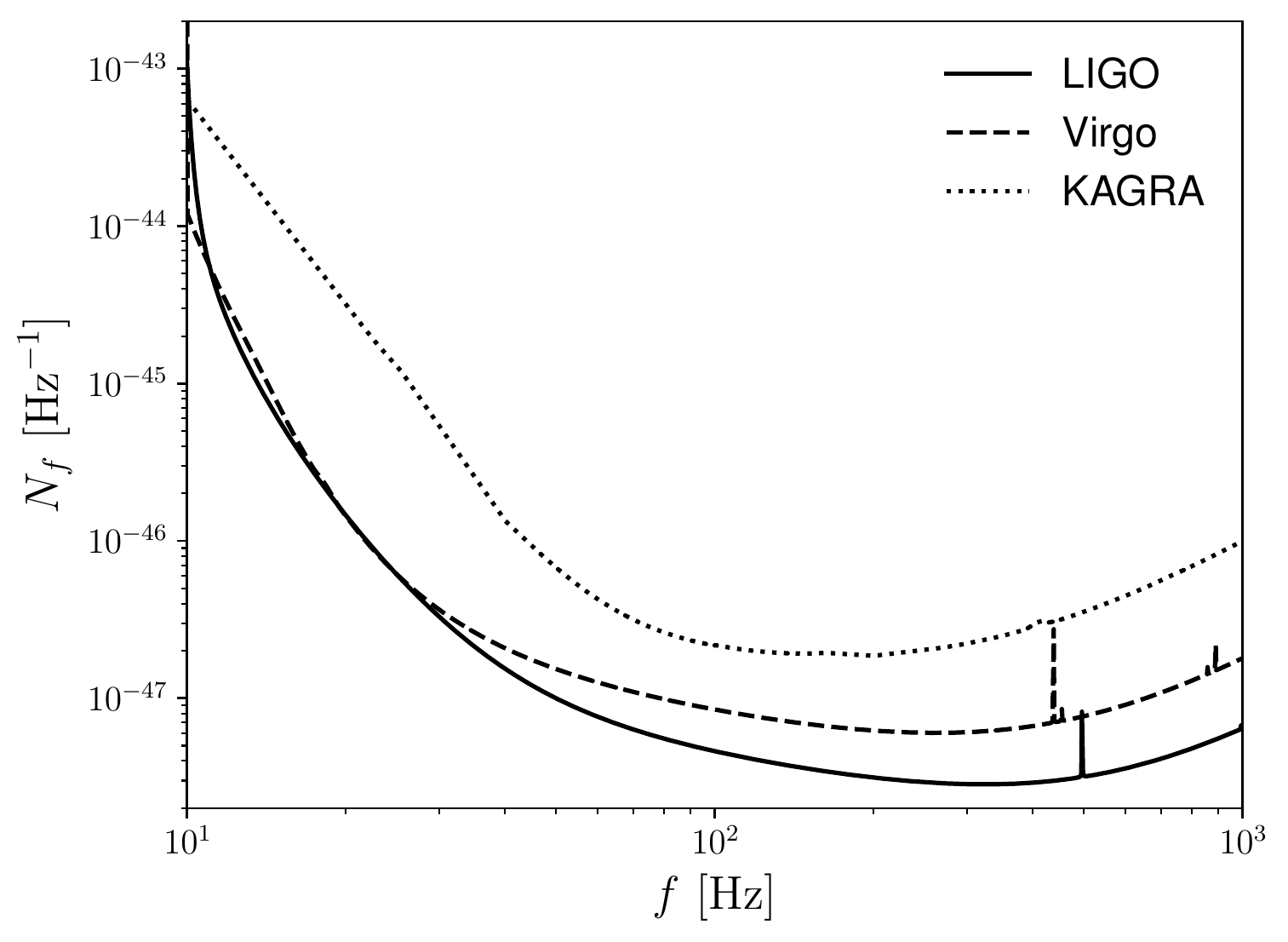}
              \caption{The sensitivity curves for the various detectors which constitute the XGN. For the two \ligo{} detectors we consider the $A+$ advanced \ligo{} design sensitivity curve, for  \virgo{} the O5-High curve, and for \kagra{} the sensitivity corresponding to a horizon at 128 Mpc. All curves are taken from \cite{LIGOPub}.} \label{fig:s_h_ligo}
            \end{center}
          \end{figure}
          \begin{figure*}
            \begin{center}
              \includegraphics[width=0.7\textwidth]{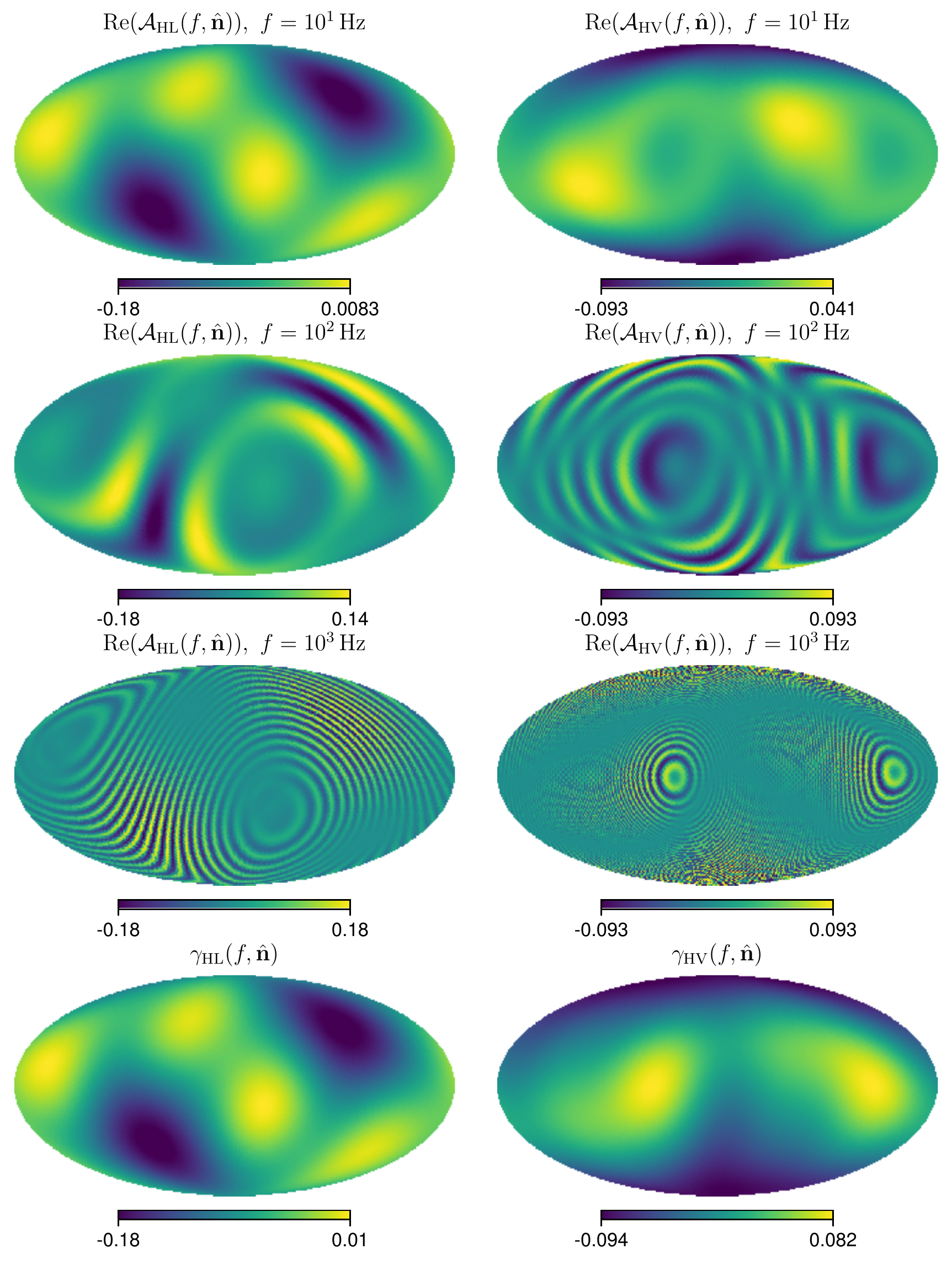}
              \caption{Antenna pattern $\resp_{AB}$ for the Hanford-Livingston (HL) and Hanford-\virgo{}  (HV) baselines of the XGN on different frequencies (left and right columns respectively). Larger frequencies allow the network to probe increasingly smaller scales. The bottom panels show the corresponding overlap functions $\gamma^I(f,\nv)$, defined in Eq. \,(\ref{eq:gamma_I}).}
             \label{fig:LIGO_antenna}
           \end{center}
          \end{figure*}
          \begin{figure*}
            \begin{center}
              \includegraphics[width=0.4\textwidth]{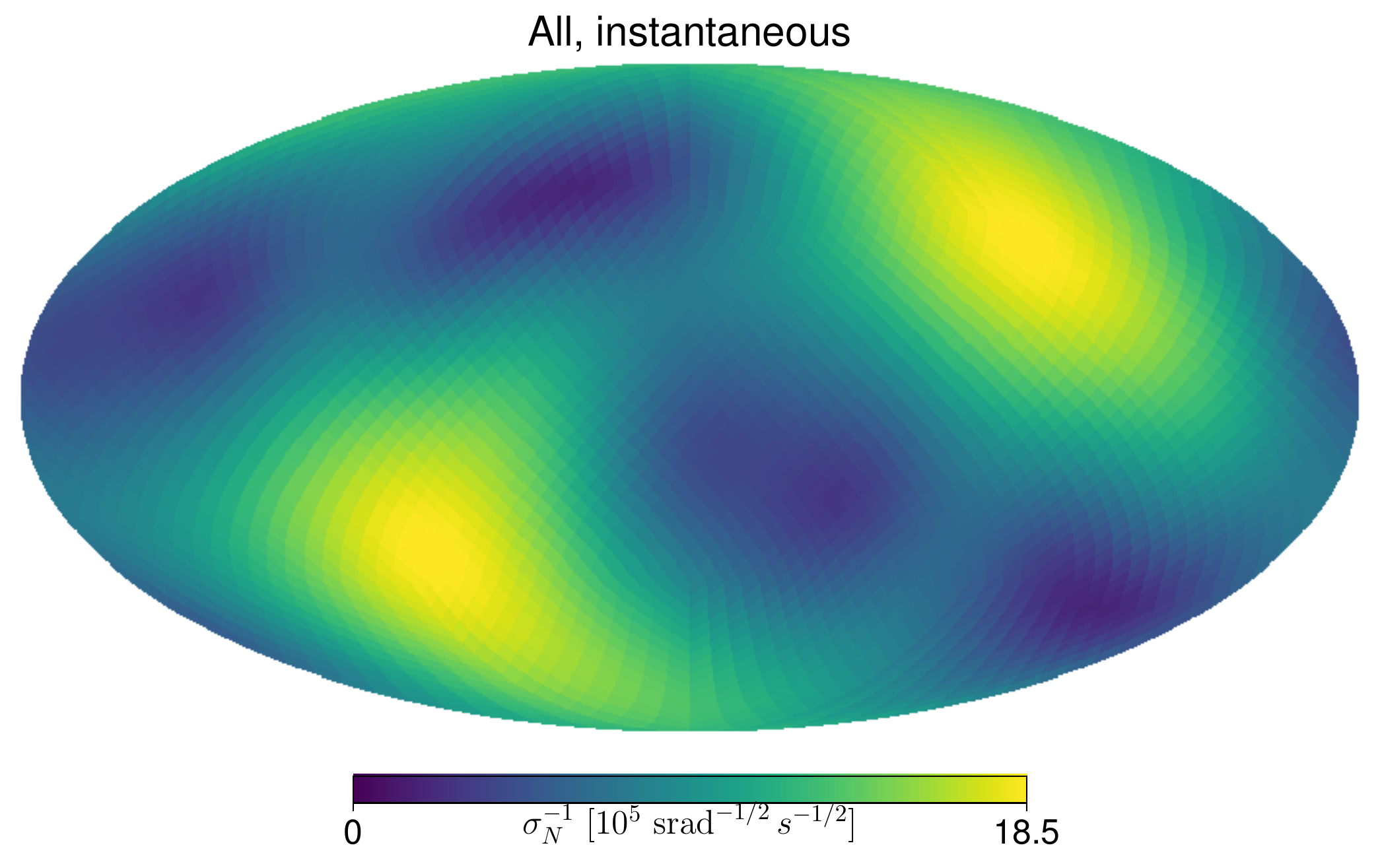}
              \includegraphics[width=0.4\textwidth]{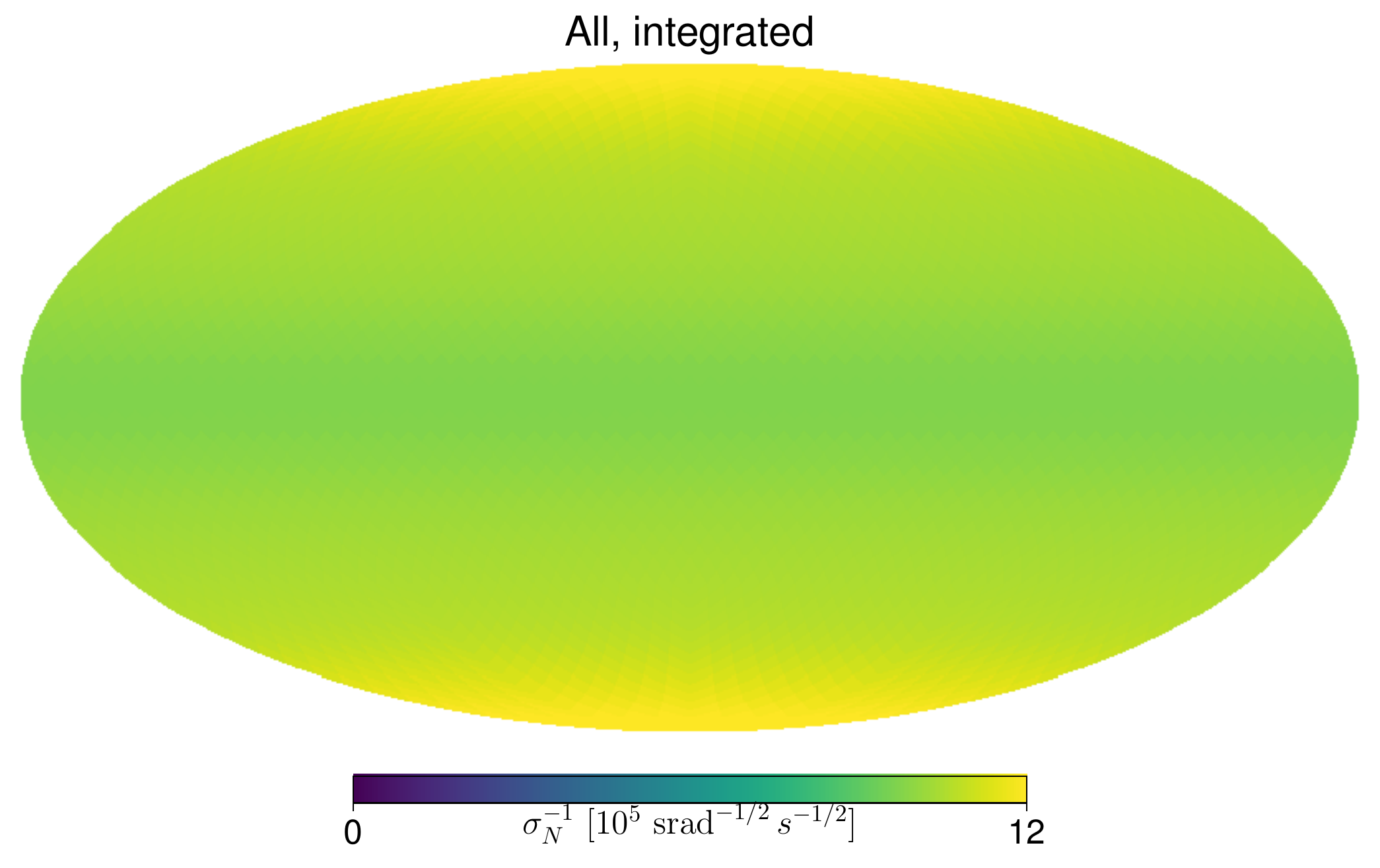}
              \caption{Square root of the map-level inverse noise variance for the full XGN across the sky. Results are shown for a single timeframe (left) and for a full year of observation (right).}
              \label{fig:LIGO_noivar}
            \end{center}
          \end{figure*}
          The detector noise properties have a number of contributions -- seismic, thermal and quantum -- which lead to a non-trivial frequency dependence. We obtain measured and forecast noise PSDs for the different detectors from  \cite{LIGOPub}.  We plot the $N_f$ for the different detectors in Fig \ref{fig:s_h_ligo}. For the Hanford and Livingston detectors, we use the Advanced LIGO {\sl design} curve \citep{designcurve}, and we assume they have essentially equivalent sensitivities. Under this assumption, they are the most competitive across thw whole frequency range. The \virgo{} sensitivity is comparable to the US detectors albeit with somewhat a poorer sensitivity at high frequencies. \kagra{}, as currently envisioned has poorer sensitivity as compared to \ligo{} and \virgo{}. All detectors are assumed to be uncorrelated. The PSD curves used here, from \cite{LIGOPub},  are based on forecasts and may differ from the actual achieved sensitivities of these experiments. We have however verified that we our implementation is able to reproduce the sensitivities to the GWB monopole presented in \cite{LIGOScientific:2019vic}, and the anisotropy noise power spectrum presented in \cite{Renzini:2019vmt} for the \ligo{} O1 and O2 runs using the corresponding sensitivity curves.

       \subsubsection{Map noise properties}\label{sssec:Examples.aLIGO.map}
         As in the case of \lisa{}, we start by inspecting the antenna patterns and inverse noise variance maps of the experiment. The antenna patterns for the Hanford-Livingston and the Hanford-\virgo{} baselines are shown in Figure \ref{fig:LIGO_antenna}, together with their overlap functions (bottom row of the same figure), to which the antenna patterns converge at low frequencies ($f\,{\bf b}_{AB}\ll c$). The corresponding instantaneous and 1-day cumulative inverse noise variance maps for the Hanford-Livingston baseline and for the full network are shown in Figure \ref{fig:LIGO_noivar}. After a full period, the sky coverage achieved by the network is roughly homogeneous, with $\sim10\%$ level variations in noise variance across the sky. All maps are shown in equatorial coordinates.

\begin{figure}
           \begin{center}
             \includegraphics[width=0.47\textwidth]{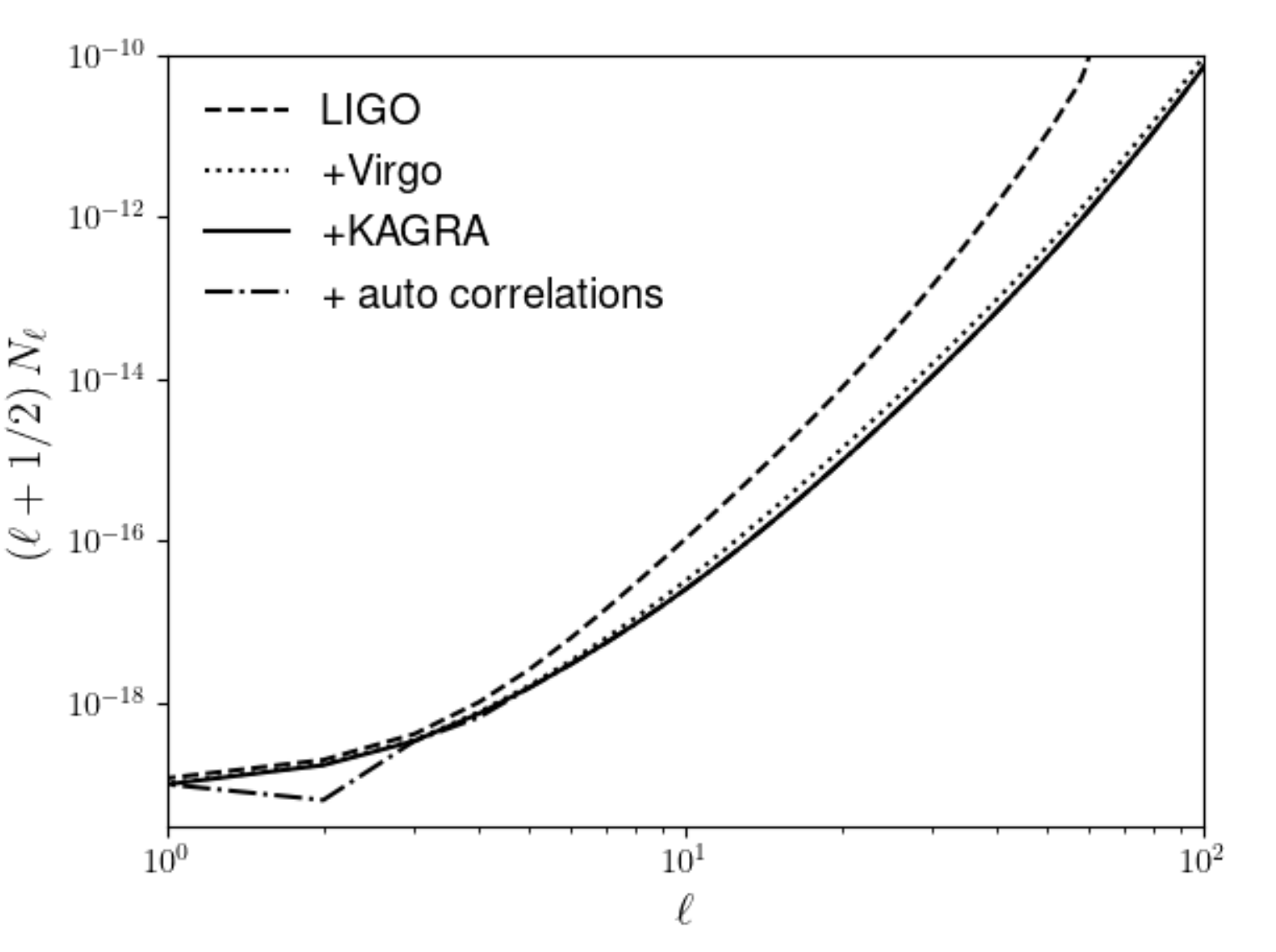}
             \caption{The noise power spectra for three different configurations of the XGB (solid, dotted and dashed lines), including only cross-correlations between detectors. The addition of auto-correlations (dot-dashed), improves the measurement of the $\ell=2$ and 4 modes. For comparison, the current estimate of the anisotropic astrophysical background power spectrum,  Eq.\,(\ref{CellOmega}), has an  amplitude $(\ell+1/2)C_{\ell}\propto10^{-25}$ for the reference frequency $f_{\text{ref}}=63$ Hz \cite{Cusin:2018avf, Cusinnew}.}\label{fig:nell_ligo} 
           \end{center}
         \end{figure}

         In Fig. \ref{fig:nell_ligo} we plot, in dashed, dotted and solid lines, the angular noise power spectra for three possible configurations of the XGN, using solely the cross-correlation between the various instruments. We choose a pivot frequency $f_{\text{ref}}=63$ Hz. For the Hanford-Livingston baseline, the angular resolution poor, and the noise power spectrum raises from $N_{\ell=2}\sim10^{-19}$ by almost three orders of magnitude at $\ell=10$, corresponding to an angular scale of $\theta\sim20^\circ$. The addition of the \virgo{} and \kagra{} baselines improves the sensitivity on smaller scales (about a factor $\sim4$ at $\ell=10$). 

         The noise properties of ground-based detectors are complicated (e.g. non-stationary, non-Gaussian), and modelling them accurately enough to be able to use auto-correlation data to build GWB intensity maps is probably unfeasible. In spite of that, it is instructive to explore what additional information those auto-correlations would bring. As shown in Fig. \ref{fig:LISA_Nell} for the case of LISA, the angular dependence of the antenna pattern for auto correlations is concentrated on low, even multipoles $\ell\lesssim6$. This is even more so for ground-based detectors with a trivial transfer function ${\cal T}=1$. The angular noise power spectrum for the XGN after adding the auto-correlations of all detectors is shown as a dot-dashed line in Fig. \ref{fig:nell_ligo}. As expected, auto-correlations are mostly able to improve the sensitivity to the $\ell=2$ and $\ell=4$ modes, for which $N_\ell$ improves by a factor $\sim10$ and 2 respectively. At the reference frequency used here ($f_{\rm ref}=63\,{\rm Hz}$), the expected amplitude of the astrophysical anisotropies is $(\ell+1/2)C_\ell\sim10^{-25}$ \citep{Cusin:2018avf, Cusinnew}, approximately four orders of magnitude lower than the noise spectrum.

     \subsection {Einstein Telescope} \label{ssec:Examples.ET}
       \begin{figure}
         \begin{center}
           \includegraphics[width=0.47\textwidth]{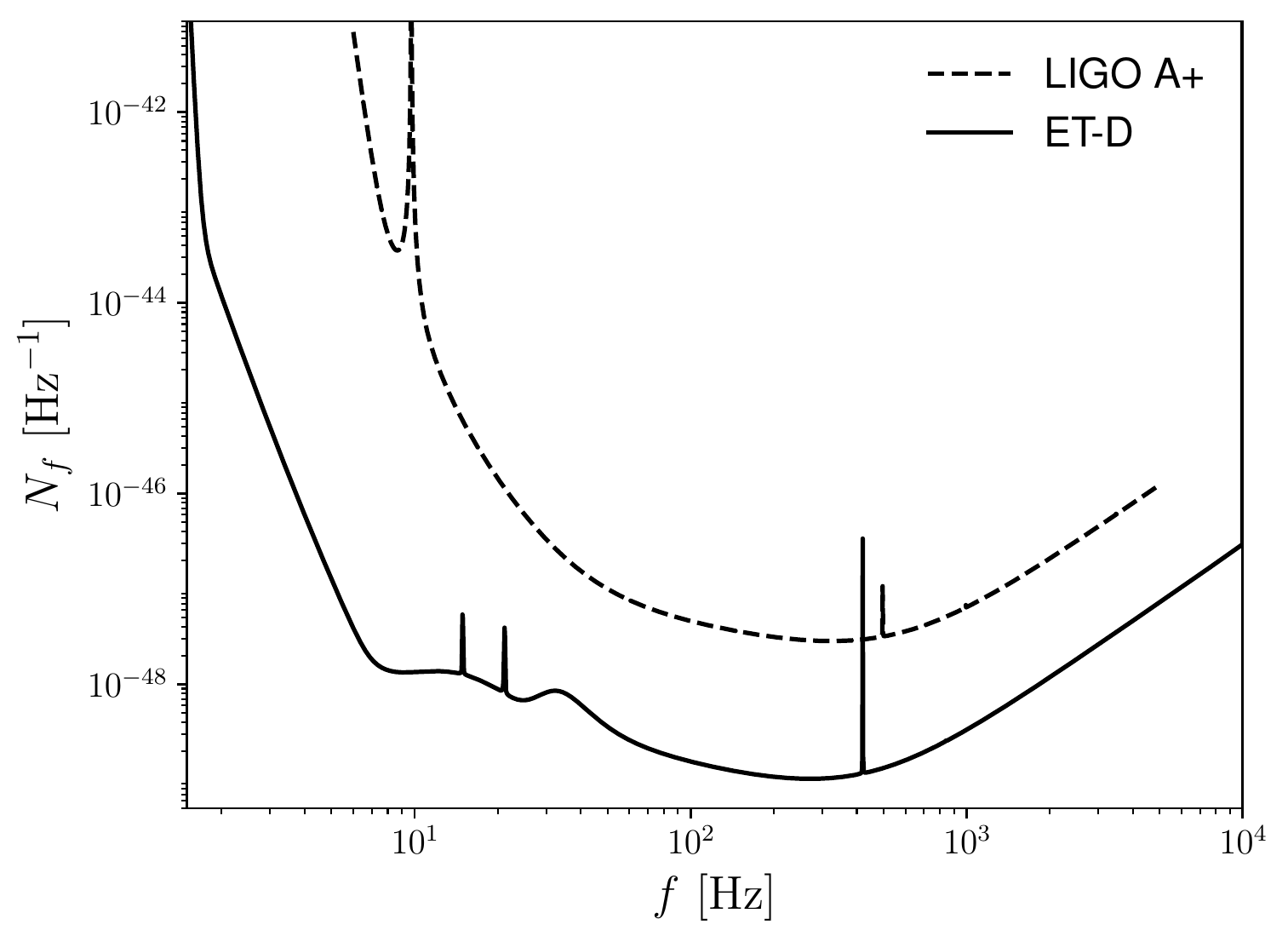}
           \caption{Power spectral density of the ET-D noise $N_f$ \cite{ETPub} compared to the $A+$ design sensitivity curve of \ligo{} \cite{LIGOPub}.}\label{fig:SET}
         \end{center}
       \end{figure}
       \begin{figure}
         \begin{center}
           \includegraphics[width=0.47\textwidth]{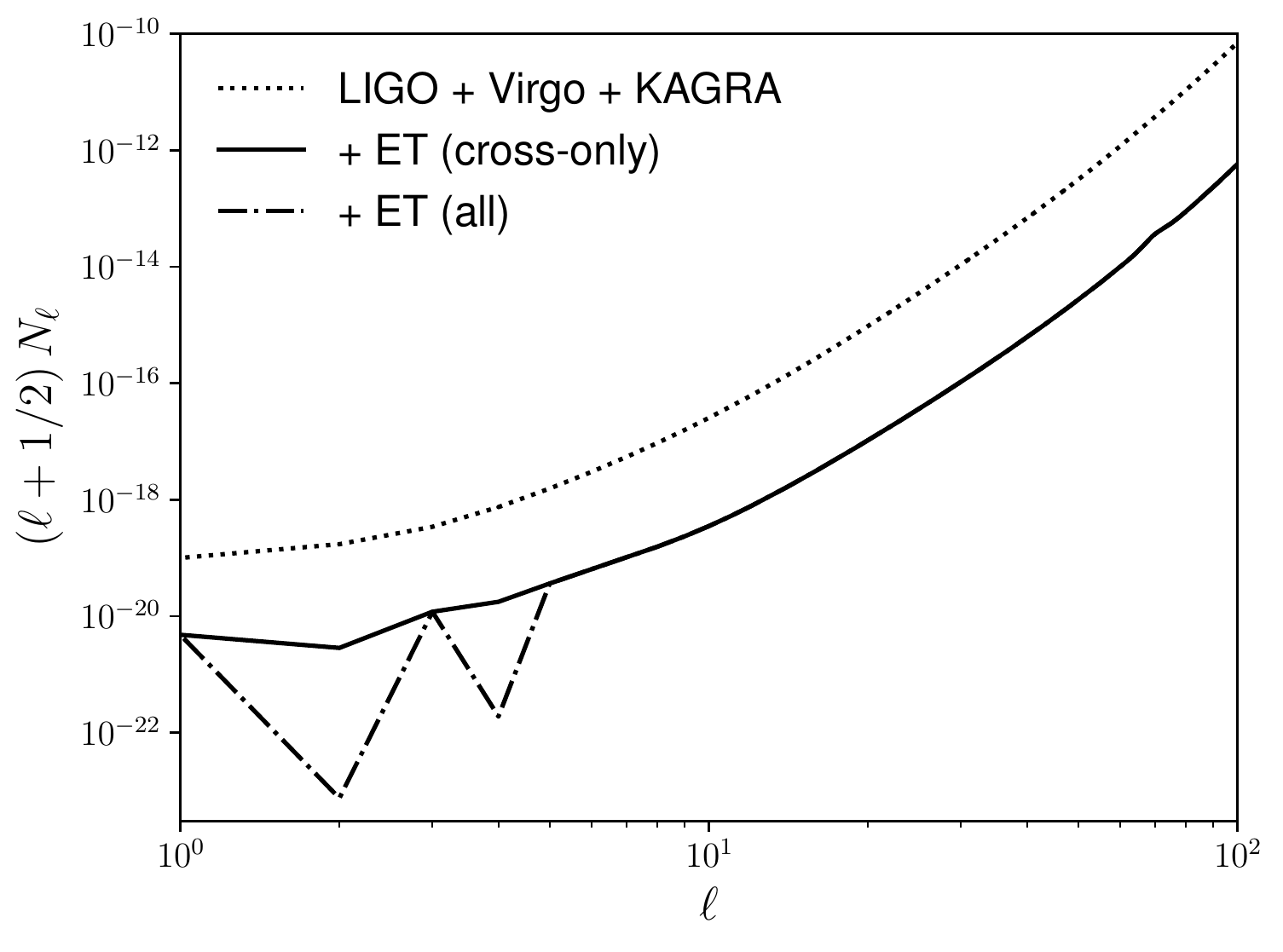}
           \caption{The noise spectrum for the Hanford-Livingston-\virgo{}-\kagra{} configuration using only cross-correlations and including autocorrelations, for a reference frequency $f_{\text{ref}}=63$ Hz.}\label{fig:ET_LIGO}
         \end{center}
       \end{figure}
       The Einstein Telescope\footnote{\url{http://www.et-gw.eu/}} (ET, \citep{Hild:2008ng}) is a hypothetical, third generation, ground based interferometer consisting of three 10-kilometre arms arranged in an equilateral triangle with two detectors at each node (allowing it to measure polarization of the incoming gravitational waves). The significantly larger sensitivity of ET with respect to the other ground-based detectors considered in the previous section makes exploring the benefits of its addition to the global network of gravitational wave detectors an instructive exercise.
       
       The site for ET has not yet been chosen, with several locations still under consideration. We arbitrarily choose a location in Sardinia, close to one of the surveyed sites (the specific coordinates and orientation angle of the triangular network are shown in Table \ref{tab:coords}).

       The ET system will consist of a low frequency instrument (which covers $1$ to few $\times 10^2$ Hz) and a high frequency instrument the spanning the $10^1-10^4\,{\rm Hz}$ range. To reduce gravity gradient noise and seismic noise, and to extend significantly the sensitivity toward low frequencies, ET will be built a few hundred meters underground. Here, we use the sensitivity curve of the instrument in the so-called D-configuration \cite{Hild_2011}. The expected improvement in strain sensitivity with respect to the Advanced \ligo{} design sensitivity is expected to be around a factor $10$ (i.e a factor $\sim100$ in $N_f$), as shown in Figure \ref{fig:SET}. For simplicity, we will assume the noise in ET is 20\% correlated between any two different detectors.

       Since the use of auto-correlations to produce GWB intensity maps may be challenging for ground-based detectors, we have explored two different configurations for a future global network including both XGN and ET. First, since the three detectors in the ET triangle will have correlated noise properties, we consider the case in which ET is only ever cross-correlated with other detectors (i.e. all auto- and cross-correlations between ET detectors are discarded). Given the dependence of $N_\ell$ on the product of detector PSDs, we would expect the resulting angular noise power spectrum to decrease by a factor ${\cal O}(100)$ with respect to the XGN-only case. This is indeed the case, as shown in Figure \ref{fig:ET_LIGO}, which displays, as dotted and solid lines, the angular noise power spectra for the XGN with and without the addition of cross-correlations with ET respectively. Secondly, we explore the benefits of adding all of the ET internal auto- and cross-correlations. Using the same arguments, we would expect these to improve the sensitivity by an additional two orders of magnitude on the scales they are sensitive to. The resulting noise curve is shown as a dot-dashed line in Fig. \ref{fig:ET_LIGO}. Given the small size of the ET arms when compared with the wavelengths it is most sensitive to (its transfer frequency is $f_*\simeq4\,{\rm kHz}$), ET itself mostly reduces the noise in the $\ell=0$, 2 and 4 multipoles, for which $N_\ell$ sees an improvement of roughly an additional two orders of magnitude.

  \section{Discussion}\label{sec:Discussion}
    \begin{figure}
      \begin{center}
        \includegraphics[width=0.47\textwidth]{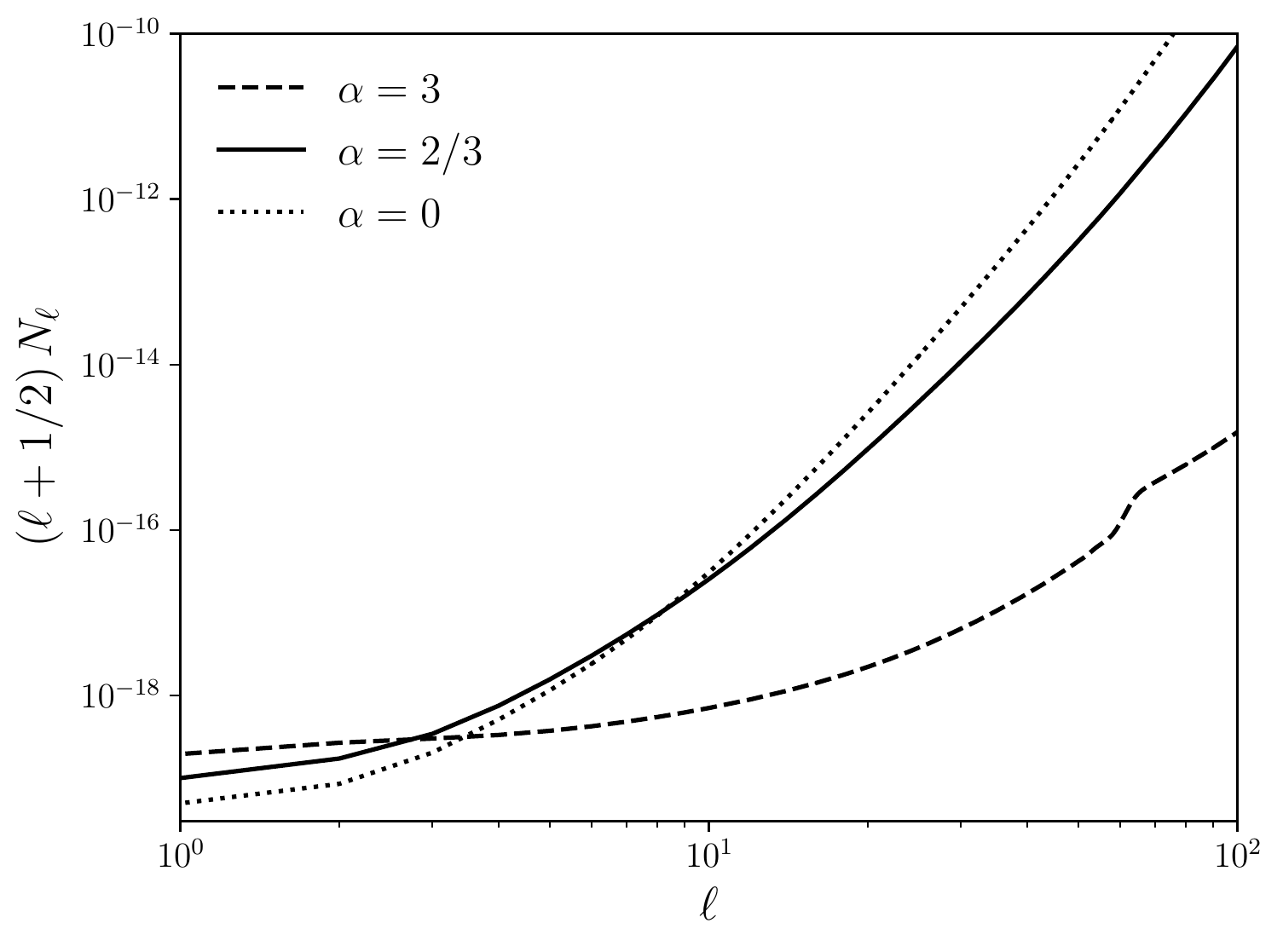}
        \caption{The noise angular power spectrum for the XGN with different values of the spectral index $\alpha$ at a reference frequency $f_{\text{ref}}=63$ Hz.}
        \label{fig:alphas} 
      \end{center}
    \end{figure}
    In this paper we have constructed a model of the angular power spectrum of the noise which will arise in searches for anisotropic gravitational wave backgrounds by networks of interferometers. The formalism we have developed is general and depends on a number of different factors: the noise spectral density of the time series, the network configuration, and its motion. It is, as we have shown, possible to put all these together to find an estimate of the noise imprinted on the sky by the apparatus and compress it into a noise per multipole $N_\ell$ that can be used to produce forecasts for the detectability of different cosmological or astrophysical signals.

    We have applied our formalism to three existing or planned GW experiments: the four operating ground-based detectors combined into an ``extended ground-based network'' XGN, the combination of the XGN and the future Einstein Telescope, and the \lisa{} satellite network. The estimated $N_\ell$s for an astrophysical background of binaries are shown in Fig. \ref{fig:LISA_Nell} for \lisa{} at a reference frequency of 0.01 Hz, and for the different ground-based experiments in Figures \ref{fig:nell_ligo} and \ref{fig:ET_LIGO} at $f=63\,{\rm Hz}$.
    
    In the case of \lisa{}, we find that the instrument will mostly be sensitive to the multipoles $\ell=0$, 2 and 4. The sensitivity to odd multipoles is reduced due to the parity of the antenna patterns, and the $N_\ell$ grows rapidly after $\ell=4$. We have also shown that most of the sensitivity is contained in the so-called ``A'' and ``E'' TDI channels, with the T channel (also called the Sagnac signal) being mostly dominated by noise. In the case of the XGN, the long baselines compared to the gravitational wave wavelengths break the parity of the antenna patterns and enable the measurement of both odd and even multipoles. The angular resolution of the network however is still somewhat limited to multipoles $\ell\lesssim10$. We have shown that, in both cases, after a full rotation of the networks, they are able to sweep the celestial sphere rather homogeneously, with relatively small noise variations across the sky.
    \begin{figure}
      \begin{center}
      \includegraphics[width=0.51\textwidth]{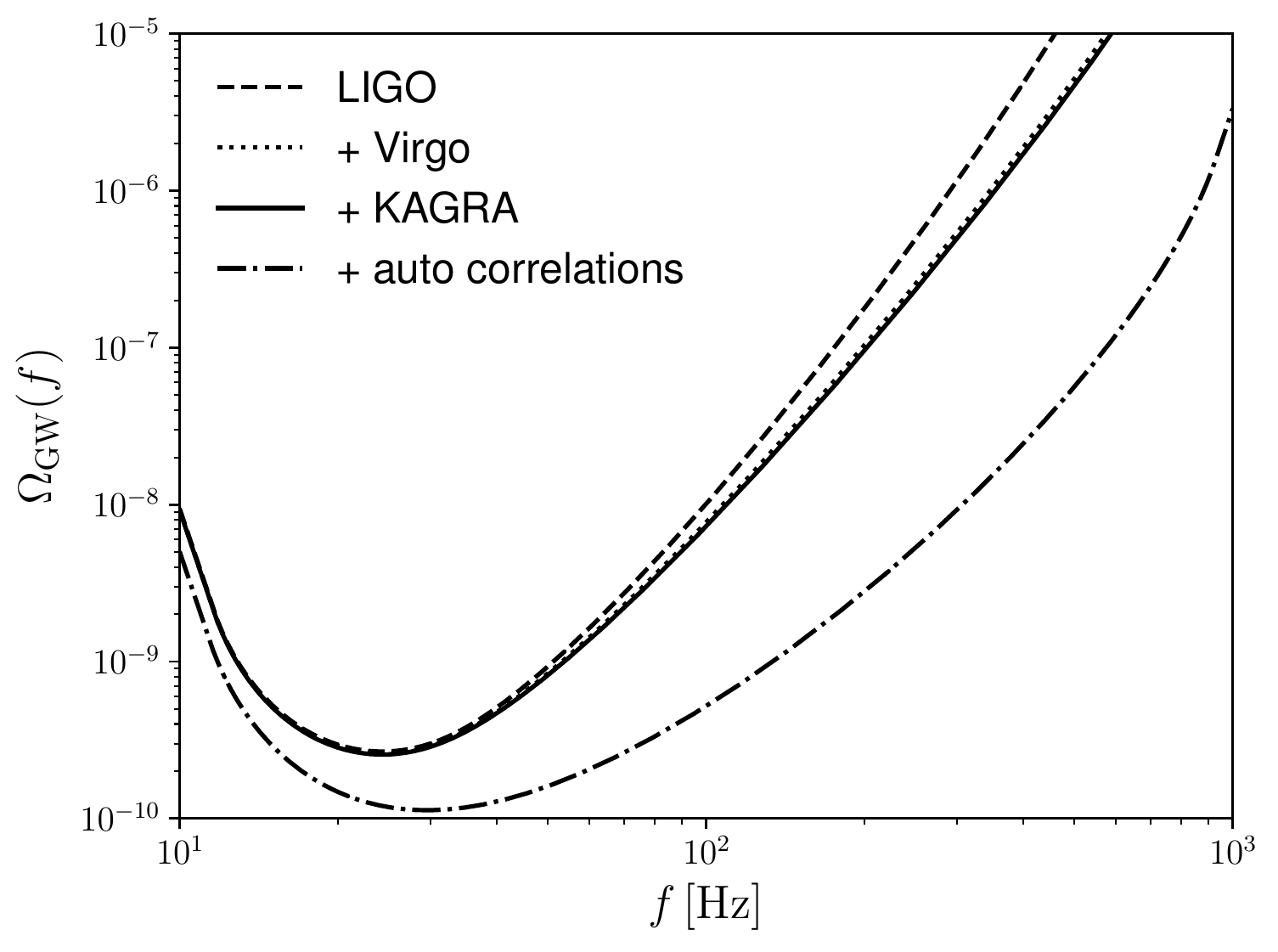}
      \caption{Monopole power-law integrated sensitivity curve \citep{Thrane:2013oya} for different configurations of the XGN.} \label{fig:monopole}
      \end{center}
   \end{figure}
    
    Our treatment has focused on the search of gravitational waves with a factorisable frequency dependence.
    In particular, we have presented results for astrophysical GWBs arising from binary mergers, since we expect this to be the dominant signal (see e.g. \cite{Contaldi:2016koz,Cusin:2017mjm,Cusin:2017fwz, Jenkins:2018uac, Cusin:2018avf, Cusin:2018rsq,Cusinnew,Cusin:2019jhg,Pitrou:2019rjz, Alonso:2020mva}). This means that, in our choice of ${\cal E}_f$ we have chosen a power law with spectral index $\alpha_I=-7/3$, corresponding  to a slope in energy density $\Omega_{\text{GW}}\propto f^{\alpha}$ with $\alpha=2/3=-3+\alpha_I$. It is however instructive to consider what the noise angular power spectrum will look like for different choices of $\alpha$. Figure \ref{fig:alphas} shows the predicted noise for the XGN assuming a flat energy spectrum $\alpha=0$, corresponding to the typical scaling of cosmological backgrounds, and a flat intensity spectrum $\alpha=3$. The shape of the $N_\ell$ therefore may depend crucially on the spectrum of the signal being searched for, as well as on the reference frequency at which the spectrum is evaluated. Both factors must be taken into account when interpreting these noise curves.    

    Although our discussion is mainly on the detectability of GWB anisotropies, studies of the GWB have mostly focused on the monopole, i.e. the sky-averaged mean, of the background. This is expected to be 4-5 orders of magnitude larger than any anisotropic signal, and may be detectable with future experiments. As described in Section \ref{sssec:gws.cont.monopole}, our formalism can also be used to make predictions for the sensitivity to the monpole of a given network. Fig.\,\ref{fig:monopole} shows the power-law integrated (PI) sensitivity curves (as defined in \cite{Thrane:2013oya}) for different configurations of the XGN, in good agreement with standard results in the literature (see e.g. \cite{Renzini:2019vmt, LIGOScientific:2019vic}).

    We have implemented the formalism described here into a software package, {\tt schNell}, that we make publicly available\footnote{The source can be found in \url{https://github.com/damonge/schNell}, and its documentation, including installation instructions, can be found in \url{https://schnell.readthedocs.io}}. This code is able to provide fast estimates of the $N_\ell$ for arbitrary networks of detectors, as well as a variety of other GWB noise properties, including noise variance maps and PI curves.

    While we encourage the use of this numerical model to make accurate precitions of $N_\ell$, it is also instructive to find approximate expressions for this quantity that describe how the different factors play a role. For Earth-based detectors, an analytic model for the inverse noise for the detector pair $AB$ which is ${\cal O}(1)$ accurate for an astrophysical and cosmological background ($\alpha=2/3$ and $\alpha=0$) is given by: 
    \begin{align}\label{eq:NellAn}
      &(N_{AB})_{\ell}^{-1}=10^{23}(2\pi)^{-2\alpha}\left(\frac{N_f^{\text{LIGO}}}{N^A_f}\right)\left(\frac{N_f^{\text{LIGO}}}{N^B_f}\right)\nonumber\\
      &\quad\times \left(\frac{f_{\text{ref}}}{ 100\text{Hz}}\right)^{-2\alpha}\left(\frac{b_{AB}}{3000 \text{km}}\right)^{-2\alpha+5}\left(\frac{T_{\text{obs}}}{\text{yr}}\right)\nonumber\\
      &\quad \times \int_{\beta_{\text{min}}}^{\beta_{\text{max}}} d\beta\, \beta^{-6+2\alpha}|j_{\ell}(\beta)|^2\,,
    \end{align}
    where $\beta\equiv 2\pi b_{AB} f$ and $j_{\ell}(x)$ is the spherical Bessel function. We assumed that the noise spectral density is given by an inverse top hat function (i.e. $N_f^{-1}$ is constant in $[f_{\text{min}}, f_{\text{max}}]$ and 0 outside this interval). All quantities in (\ref{eq:NellAn}) have been normalized with respect to typical values for the two \ligo{} detectors in the designed configuration and $N_f^{\text{LIGO}}\equiv 0.8 \cdot 10^{-47}$ Hz$^{-1}$, $\beta_{\text{min}}=2\pi (b_{AB}/ 3000 \text{km})(f_{\text{min}}/(100\text{Hz}))$ and analogous for $\beta_{\text{max}}$. A key simplifying assumption in Eq.\,(\ref{eq:NellAn}) has been to discard any angular dependence in the overlap function. For a fixed $\alpha$ (i.e. a fixed background), Eq.\,(\ref{eq:NellAn}) depends on different parameters: the observation time $T_{\text{obs}}$, the baseline $b_{AB}$ between detectors $A$ and $B$ and the features of the  sensitivity of the two detectors $AB$, i.e. the values of the plateaux $N_A$ and $N_B$, $f_{\text{min}}$ and $f_{\text{max}}$. Unfortunately we have not been able to find a similar approximate formula for \lisa{}. This should however not be a significant obstacle in producing forecasts for this experiment since, as described in Section \ref{ssec:Examples.LISA}, the $N_\ell$ is in practice characterized by three numbers: the values at $\ell=0$, 2 and 4.
   
    \begin{figure}
      \begin{center}
        \includegraphics[width=0.5\textwidth]{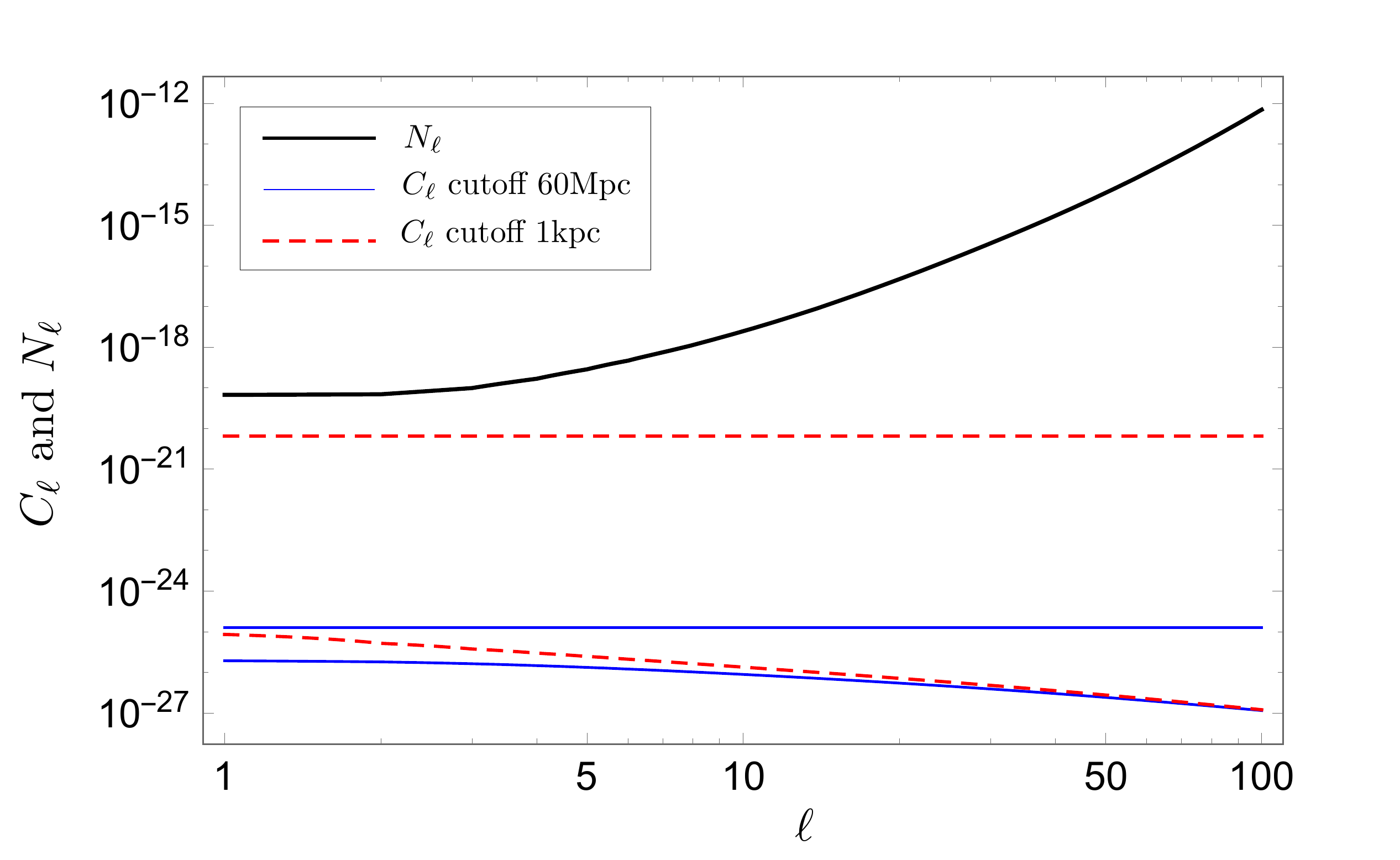}
        \caption{The noise spectrum for the Hanford-Livingston-\virgo{}-\kagra{} and $\alpha=2/3$ compared to the signal expected from a background of black hole mergers at $f=63$ Hz. We choose two different values for the threshold of detection of individual sources: 60 Mpc (solid thick line) and 1 kpc (red dashed line). For each choice we plot the contribution from clustering ($\propto 1/\ell$) and the contribution from Poisson noise  (flat off-set) \cite{Cusinnew, Alonso:2020mva}.} \label{LIGOSignal}
      \end{center}
    \end{figure}
    Our predictions for the angular power spectrum of current and future experiments can be readily compared with theoretical predictions for different types of GWBs. Predictions for the astrophysical background from binary mergers found in the literature (e.g. \cite{Cusin:2018avf, Cusin:2018rsq, Cusin:2019jhg, Cusinnew}) place the expected signal in the range $(\ell+1/2) C_{\ell} \sim 10^{-25}$ at $f_{\rm ref}=63$ Hz, and $(\ell+1/2) C_{\ell} \sim10^{-30}$ at $f=0.01$ Hz. This is between 4 and 6 orders of magnitude lower than the expected noise spectrum for the XGN and \lisa{} respectively, which would reduce to $\sim2$ orders of magnitude at $\ell=2$ if the Einstein Telescope was added to the XGN (see Figs. \ref{fig:LISA_Nell}, \ref{fig:nell_ligo}, and \ref{fig:ET_LIGO}). Based on existing or projected sensitivities and astrophysical models, it therefore seems unlikely that we will be able to detect the anisotropies of a gravitational-wave background with a simple spherical harmonic decomposition approach without major improvement of the detector designs or a substantial change in the detection techniques. We can however ask ourselves what it would take to achieve a detection of the GWB anisotropies. Although $N_\ell$ scales steeply with the inverse of the strain noise to the fourth power, this would still require an improvement of the sensitivity of the XGN instruments \cite{designcurve} by a factor of $\sim30-50$ in order to achieve an $N_\ell$ comparable with the signal one might expect from the astrophysical background on large scales. The prospect for cosmological backgrounds \cite{Caprini:2018mtu,Geller:2018mwu} is probably more dire. With negligible primary anisotropies, the typical size of the fluctuations in intensity with respect to the monopole is $\sim 10^{-5}$. This means that in a given frequency band the size of the monopole has to be $10^{6}$ times above the noise plateau to have detections of first multipoles (ignoring cosmic variance contributions) with SNR of 10.

    As was shown in \cite{Jenkins:2019uzp, Cusinnew, Alonso:2020mva}, in the frequency band of ground-based interferometers, the astrophysical background is dominated by a ``pop-corn'' component associated to the discreteness in time of binary merger events, which has a larger amplitude than the clustering component and depend on the horizon for the detectability of resolvable events (see \cite{Cusinnew,Alonso:2020mva}). Figure \ref{LIGOSignal} shows these two components of the power spectrum for the astrophysical GWB for two different distance cutoffs $d$ (i.e. the distance up to which all events are detected and removed from the data) in dashed red ($d=1\,{\rm kpc}$) and solid blue ($d=60\,{\rm Mpc}$). For each pair of curves, the flat component with a higher amplitude corresponds to this pop-corn contribution, while the sub-dominant curve shows the smooth clustering component. One might argue that the pop-corn contribution is a more viable target for anisotropic GWB searches, given its higher amplitude. While this would be interesting prospect as a first detection, it would effectively only provide information on a single number, and therefore little astrophysical information  (i.e. about the source properties, merger rates, etc.) would be gleaned from this measurement, not adding much to a detection of the monopole of the same background component or to direct measurements of resolved sources. It is worth recalling that, in the mHz \lisa{} band, the only contribution to the anisotropy comes from clustering and no popcorn component is present for a background of stellar mass back holes.

    While the future of GWB anisotropy detection seems bleak there are a number of efforts aimed at integrating interferometric data more efficiently in order to access better constraints and defeat the noise levels with detector-specific recipes. One possibility explored by the \ligo{}-\virgo{} collaboration is to perform narrow sky searches with a radiometer method~\cite{Ballmer2006} informed by the light matter distribution. This approach would not be able to measure the anisotropy over the whole sky, but, if successful it would provide a localised measurement of the GWB. Cross-correlated searches with large scale structure are expected to improve the SNR of auto-correlation by typically a factor 10 \citep{Alonso:2020mva}.  The potential of cross-correlation, while not sufficient to get a detection of the anisotropy for  present and planned instrumental noise levels, should be kept in mind when looking at new detection strategies. It is clear that more new ideas and techniques need to be brought to the fore if we are to effectively measure anisotropies in the gravitational wave background. 

  \section*{Acknowledgements}
    We are extremely grateful for discussions with Neil Cornish, Vuk Mandic, Andrew Matas, Ioannis Michaloliakos, Cyril Pitrou, Joseph Romano, Jean-Philippe Uzan, and Bernard Whiting. DA acknowledges support from Science and Technology Facilities Council through an Ernest Rutherford Fellowship, grant reference ST/P004474/1. PGF and DA acknowledge support from the  Beecroft Trust. This  project  has  received  funding  from  the European  Research  Council (ERC) under  the  European  Union's Horizon  2020  research  and  innovation  programme  (grant  agreement No 693024)  and from the Swiss National Science Foundation.

\bibliographystyle{apsrev4-1}
\bibliography{main}

\end{document}